\newif\ifprintfig
\printfigtrue

\documentclass{emulateapj}
\usepackage{graphicx}

\newcommand\pc{{\rm\,pc}}

\newcommand\kpc{{\rm\,kpc}}
\newcommand\Mpc{{\rm\,Mpc}}

\newcommand\Gyr{{\rm\,Gyr}}

\newcommand\kmsec{{\rm\,km\,s^{-1}}}
\newcommand\kms{\kmsec}

\newcommand\msun{{\rm\,M_\odot}}
\newcommand\Msun{{\rm\,M_\odot}}

\newcommand\clock{\count0=\time \divide\count0 by 60
     \count1=\count0 \multiply\count1 by -60 \advance\count1 by \time
     \number\count0:\ifnum\count1<10{0\number\count1}\else\number\count1\fi}

\shortauthors{Dalcanton \& Stilp}
\shorttitle{Pressure in Galaxy Disks}

\begin{document}

\title{Pressure Support in Galaxy Disks: Impact on Rotation Curves and Dark Matter Density Profiles}

\author{Julianne J. Dalcanton\altaffilmark{1},
Adrienne Stilp\altaffilmark{1}
}

\altaffiltext{1}{Department of Astronomy, Box 351580, University of Washington, Seattle, WA 98195; jd@astro.washington.edu; stilp@astro.washington.edu}
  
\begin{abstract}
  Rotation curves constrain a galaxy's underlying mass density
  profile, under the assumption that the observed rotation produces a
  centripetal force that exactly balances the inward force of gravity.
  However, most rotation curves are measured using emission lines from
  gas, which can experience additional forces due to pressure.  In
  realistic galaxy disks, the gas pressure declines with radius,
  providing additional radial support to the disk. The measured tangential
  rotation speed will therefore tend to lag the true circular velocity
  of a test particle.  The gas pressure is dominated by turbulence,
  and we evaluate its likely amplitude from recent estimates of the
  gas velocity dispersion and surface density.  We show that where the
  amplitude of the rotation curve is comparable to the characteristic
  velocities of the interstellar turbulence, pressure support may lead
  to underestimates of the mass density of the underlying dark matter
  halo and the inner slope of its density profile.  These effects may
  be significant for galaxies with rotation speeds $\lesssim75\kms$,
  but are unlikely to be significant in higher mass galaxies.  We find
  that pressure support can be sustained over long timescales, because
  any reduction in support due to the conversion of gas into stars is
  compensated for by an inward flow of gas.  However, we point to many
  uncertainties in assessing the importance of pressure support in
  real or simulated galaxies.  Thus, while pressure support may help
  to alleviate possible tensions between rotation curve observations
  and $\Lambda$CDM on kiloparsec scales, it should not be viewed as a
  definitive solution at this time.

\end{abstract}
\keywords{cosmology: dark matter --- ISM: kinematics and dynamics --- galaxies: ISM --- galaxies: kinematics and dynamics -- galaxies: dwarf}

\vfill
\clearpage

\section{Introduction}  \label{introsec}

Galaxy rotation curves offer some of the strongest evidence for dark
matter on kiloparsec scales \citep[][and references therein]{sofue01}.
This evidence rests on the assumption that rotating disk galaxies are
in equilibrium, such that the outward centripal force of the rotating
disk exactly balances the inward gravitational force from the mass
interior to the disk.  The rotation speed seen in atomic gas at large
radii therefore can constrain a galaxy's mass.  The apparent
``excess'' mass inferred from the rotation curve, compared to the much
smaller mass inferred from observations of the baryons within the
galaxy, is assumed to be due to dark matter.  The dark matter
furthermore must be distributed with an approximately $\rho\propto
r^{-2}$ density profile in the outer galaxy, to reproduce the flat
rotation curve that is observed.

In the centers of galaxies, constraints on the dark matter require a
more accurate accounting of the baryonic mass, since baryons make up a
larger fraction of the mass in a galaxy's inner regions
\citep[e.g.,][]{kalnajs1983,kent1986,vanalbada1986}. Unfortunately,
assigning a mass to the baryonic component of galaxy is uncertain, due
to our limited knowledge of the stellar mass-to-light ratio, the
effects of dust extinction, and the contributions of atomic and
molecular gas.

Alternatively, the difficulties in probing inner halos can be
reduced by studying only galaxies with intrinsically low baryon
fractions, such as dwarf and low surface brightness galaxies.  These
galaxies have no central bulges, low baryonic surface densities, and
high gas fractions, thus minimizing the impact of uncertainties in
the stellar mass-to-light ratio.  Detailed analyses of such galaxies
tend to show more slowly rising rotation curves than in massive
spirals.  Part of the slow rise is due to the reduced contribution of
baryons to their inner density profiles \citep[although,
see][]{swaters2009}.  However, the slowly rising inner rotation curve is
also indicative of shallow inner density profiles for the dark matter
halo \citep[$\rho\propto r^{\alpha}$, with $\alpha\lesssim-0.25$; see
the thorough review by][and references therein]{deblok2009}.  These
inner slopes are in apparent conflict with predictions from numerical
simulations of dark matter halos in $\Lambda$CDM, which favor inner
power law slopes with $\alpha$ steeper
than -0.75 on kiloparsec scales and below
\citep[][]{dubinski1991,navarro1997, moore1998, colin2004,navarro2004,
  hayashi2004,navarro2008,stadel2009}.  When observational
uncertainties in deriving rotation curves from long-slit spectra are
taken into account, the apparent conflict is weaker
\citep[][]{spekkens2005,rhee2004, hayashi2004,valenzuela2007},
although the discrepancies between observations and theory still
appear to persist when full velocity fields are considered
\citep[e.g.,][]{spano2008, kuzio2008, kuzio2009, deblok2008}.


In addition to the possible discrepancy in the inner slope of the
density profile, the overall density likewise appears to be
significantly lower than seen in simulated halos \citep{alam2002}.
The tension between the observed and predicted halo densities has been
somewhat reduced by the latest WMAP results \citep{maccio2008}, but
significant offsets remain \citep[see the discussion
in][]{sellwood2009}.

Taken together, the rotation curve analyses are some of the strongest
lines of evidence that the success of $\Lambda$CDM on large scales
breaks down when observed on kiloparsec scales.  The majority
of these analyses rely on the assumption that the galaxy's radial
equilibrium is entirely governed by the competition between gravity
and centripetal force.  However, most rotation speeds are measured
using H{\sc i} or H{\sc ii} emission lines.  These lines are emitted
by gaseous disks, which can also experience radial forces due to
pressure gradients.  These pressure gradients have the potential to
provide an additional outward force, which helps to support the disk
radially.  This support allows the disk to remain in equilibrium while
rotating more slowly than the true circular velocity associated with
the gravitational potential.  Thus, if pressure support is neglected,
then the inferred dynamical mass of the galaxy will be less than the
true mass.

Pressure support in the ISM has several possible sources.  The most
obvious is thermal support (i.e. $P_{therm}=nkT$).  However, analysis
of the temperatures and densities of the multiphase ISM suggest that
thermal pressure falls far short of what is needed to support the
observed vertical structure \citep[e.g.,][]{boulares1990}.  The second
potential source of support is magnetic.  However, it too appears to
be insufficient to support the disk \citep[e.g.,][]{deavillez2005}.
The final, and most likely dominant, source of pressure support comes
from interstellar turbulence \citep[see reviews
by][]{elmegreen2004,maclow2004}.  This turbulence is most likely
continuously driven by stellar winds and supernovae, although the
exact source is under some debate, and other driving mechanisms may
take over in regions where the star formation rate is low.  The
turbulent pressure scales with the gas density and the velocity
dispersion as $P_{turb}\sim \rho \sigma_v^2$, such that regions of
high gas density and high velocity dispersion have higher pressures.
Since both of these quantities tend to decline with radius
in galaxy disks, pressure decreases outwards, producing a pressure
gradient that can contribute to the radial support of the disk.

Corrections for pressure support have a long and varied history in the
literature.  In the early days of interferometric 21\,cm mapping of
dwarf galaxies, it was recognized that the turbulent motions within
dwarf galaxies were comparable to the amplitude of the rotation curve
\citep[e.g.,][]{tully1978}.  Thus, pressure support due to turbulent
motion had the potential to produce a significant difference between
the true circular velocity and the observed rotational speed.  It
became standard to ``correct'' the observed rotation curve for
pressure support\footnote{This correction was often erroneously
  referred to as a correction for ``axisymmetric drift'', a term that
  applies to a stellar population's difference from the true circular
  speed, due to its velocity dispersion.  The observed offset is
  produced by the different sign of orbital epicycles for stars with
  guiding centers inside and outside of the radius of observation.
  Although the mathematics is similar, the underlying physical
  situation is quite different from the pressure support of the gas.},
usually assuming either a gaussian or a constant radial profile for
the turbulent velocity \citep{tully1978,skillman1987,lo1993,
  cote2000,meurer1996}.  These corrections fell out of favor during
the last decade, presumedly as the focus turned to slightly higher
mass low surface brightness galaxies.  In addition, assessment of the
importance of pressure using more recent, higher quality data
suggested that pressure support could be neglected in most
circumstances \citep[e.g.,][]{swaters2009}.

Renewed attention on the importance of pressure support comes from the
latest generation of hydrodynamical simulations of galaxies.  Most
notably, \citet{valenzuela2007} explicitly measured the difference
between the rotational velocity of the gas and the true underlying
circular velocity, for model late-type disk galaxies with a maximum
circular velocity of $V_c\!\approx\!70\kms$.  They found that the
observed rotational velocity indeed underestimated the true circular
velocity, and argued that pressure support was responsible for a
significant portion of the difference.  Although it is likely that the
\citet{valenzuela2007} simulations were not able to fully resolve all
the relevant subgrid physics involved in generating turbulent pressure
\citep[e.g.,][]{joung2006}, their results certainly motivate a renewed
examination of the effect of pressure in disk support, particularly in
light of new observational constraints \citep[e.g.,][]{tamburro2009}.
In addition, recent simulations suggest the possibility that
estimates of pressure support based on measurements of the vertical
velocity dispersion may underestimate the true pressure support by at
least a factor of two \citep[e.g.,][]{agertz2009}.

In this paper we reassess the likely importance of pressure in
different regimes of galaxy mass, for a range of plausible galaxy
models, using modern data to constrain the distribution and kinematics
of the gas.  We first address the issue analytically, by calculating
the expected differences between the observed rotation curves and true circular
velocities, for both NFW halo profiles and realistic slowly-rising
rotation curves.  We show that the dark matter density profiles
inferred from the observed rotation curves are extremeley sensitive to
the details of the assumed pressure distribution.  Thus, small changes
in the assumptions can drastically change the scientific conclusions.
We then consider whether pressure support is a sustainable phenomena.
Gas consumption will tend to preferentially reduce the gas density in
the center of galaxies, reducing the pressure support over time.  We
calculate the evolution in the pressure support with time, and find
that inward redistribution of gas in response to decreased pressure support
tends to counteract the central gas consumption, such that pressure support
can be significant over the lifetime of a star forming disk.  We next
analyze the pressure support in a galaxy from the THINGS survey.  In
this particular case, we find that pressure support is not
significant, as we expected from the overall rotation speed of the
galaxy.  However, the exercise highlights the difficulty in accurately
assessing the amplitude of pressure support in real galaxies.
Finally, we discuss the possible limitations of application of
simplified analytic calculations of pressure support.

\section{Pressure Support for Rotation}  \label{pressuremathsec}

The circular speed of a test particle in a circular orbit is defined
as

\begin{equation}
V_c(r) \equiv \sqrt{r\,\frac{d\Phi(r)}{dr}}
\end{equation}

\noindent for an axisymmetric system, where $\Phi(r)$ is the
gravitational potential.  If the system is in equilibrium, then the
inward radial force due to gravity must be balanced by an equal
outward force.  Traditionally, one assumes that the outward force is
entirely dominated by centripetal force due to rotation with a
velocity $V_\theta$, such that

\begin{equation}
\frac{V_\theta^2(r)}{r} = \frac{G\,M(<r)}{r^2}
\end{equation}

\noindent where $M(<r)$ is the mass contained with a radius $r$, assuming that
the potential is approximately spheroidal
and thus that there is no significant gravitational force from mass 
at larger radii.  This equation implicitly assumes that all material
at a given radius is moving with the circular speed, and that that the
observed tangential velocity $V_\theta$ obeys

\begin{equation}
V_\theta(r) = V_c(r).
\end{equation}

However, in most cases $V_\theta(r)$ (the ``rotation curve'') is measured
using gas as a tracer.  Unlike ideal test particles, the gas (usually
neutral H{\sc i} or ionized H{\sc ii}) can experience an additional outward force
due to pressure.  This gas pressure can potentially provide additional radial support to
the rotating gaseous disk.  The resulting outward force allows the gas in the
disk to rotate more slowly than the circular velocity, while still
maintaining a stable circular orbit.  In such cases, interpreting the
observed angular velocity $V_\theta$ as the circular velocity $V_c$
would lead one to underestimate the mass within $r$.  

To estimate the importance of pressure support in modifying $V_\theta$, we
consider a parcel of gas at radius $r$ within a rotating disk.  We assume
that the gas has density $\rho$ within a volume $dV = dr\,dA$
(where $dA$ is the surface area of the volume normal to the radius), giving a
mass $dM=\rho\,dr\,dA$.  For a stable circular orbit, the radial
forces must be in balance, giving

\begin{equation}
0 = F_{grav} + F_{rot} + F_{P}
\end{equation}

\noindent where $F_{grav}$, $F_{rot}$, and $F_{P}$ are the forces due
to gravity, centripetal acceleration, and the pressure gradient,
respectively.

If we assume that the gravitational potential is dominated by an
approximately spheroidal mass component (as one might expect for late
type galaxies with low surface density disks), then the radial
gravitational force $F_{grav} = -\nabla_r\Phi$ is

\begin{equation}
F_{grav} = -dM \, \frac{G\,M(<r)}{r^2}.
\end{equation}

\noindent The force due to centripetal acceleration is

\begin{equation}
F_{rot} = dM \, \frac{V_\theta^2(r)}{r}
\end{equation}

\noindent and the force due to the pressure gradient is

\begin{eqnarray}
F_{P} &=& -\frac{dP}{dr}\,dr\,dA \\
      &=&  -dM \, \frac{1}{\rho(r)} \, \frac{dP(r)}{dr}.
\end{eqnarray}

Substituting back into the force equilibrium equation, cancelling $dM$, and rearranging
yields

\begin{equation}  \label{vccorrpressureeqn}
V_c^2(r) = V_\theta^2(r) - \sigma^2_r(r) \, 
                  \left[\frac{r}{P(r)}\,\frac{dP(r)}{dr}\right].
\end{equation}

To calculate the radial pressure gradient, we assume that the gas
pressure is dominated by turbulent motions, rather than thermal
processes.  One can thus approximate the local gas pressure as 

\begin{equation}  \label{pressuredefneqn}
P = \rho_{gas}\,\sigma^2 
\end{equation}

\noindent
where $\rho_{gas}$ is the density of the gas and $\sigma^2$ is the
one-dimensional velocity dispersion.  Both the gas density and
velocity dispersion tend to decline with radius, leading to a radial
gradient in the pressure, which produces an outward force.  If one
assumes that the gas disk has constant thickness
\citep[e.g.,][]{kregel2004,bottema1986}\footnote{Note that the
  galaxies in both \citet{kregel2004} and \citet{bottema1986} have
  higher average circular speeds ($V_\theta \gtrsim 100\kms$) than the
  galaxies that are relevant for this paper.  For lower mass galaxies,
  firm constraints on the radial variation in the gas scale height
  appear to be unavailable, beyond the fact that the gas distribution
  seems to much thicker (compared to the scale length) than in more
  massive galaxies \citep{roychowdhury2010}.}, then $\rho(r) \propto
\Sigma(r)$ where $\Sigma(r)$ is the mass surface density of the gas
disk.  Equation~\ref{vccorrpressureeqn} then becomes

\begin{equation}                \label{vccorreqn}
V_\theta^2(r) = V_c^2(r) \,
               \left( 1  + \frac{\sigma^2_r(r)}{V^2_c(r)} \, 
                  \left[2\,\frac{d{\rm ln}\sigma_r(r)}{d{\rm ln}r}
                           + \frac{d{\rm ln}\Sigma(r)}{d{\rm ln}r}\right]\right),
\end{equation}

\noindent where $\sigma_r$ is the radial component of the gas
velocity dispersion ellipsoid.

Equation~\ref{vccorreqn} can be used to calculate the true circular
velocity $V_c$ from the observable quantities $V_\theta$, $\Sigma$,
and $\sigma$, assuming that one can estimate $\sigma_r$ from the HI
velocity dispersion.  We will discuss this latter assumption in
Section~\ref{obslimsec} below.  Note also that the size of the
correction to $V_c$ does not depend on the absolute value of the gas
surface density, and instead only depends on its fractional rate of
change with radius.

Equation~\ref{vccorreqn} suggests that the observed tangential
velocity can differ significantly from the true circular velocity in
regions where the turbulent velocities are of the same order as the
rotational velocity.  In addition, it shows the that strength of the
pressure support depends on the fractional rate of change in the
turbulent velocities and the gas surface density with radius.
Galaxies with rapidly declining values of $\sigma_r$ or $\Sigma$ will
have more significant pressure support, for the same $\sigma_r$.

In most applications the full expressions for $\sigma_r(r)$ and
$\Sigma(r)$ can be used in equation~\ref{vccorreqn}.  However, it is
instructive to
simplify the expression by assuming that the gas surface density and
1-d turbulent velocity fall off radially like exponentials with scale
lengths $h_r$ and $h_{\sigma}$, respectively.  If so, then the force
balance equation becomes
%

\begin{equation}                 \label{expeqn}
V_c^2(r) = V_\theta^2(r) + \sigma^2_r(r) \, 
                  \left[\frac{r}{h_P}\right].
\end{equation}

where we have defined a pressure scale length $h_P$ as

\begin{equation}                 \label{hPeqn}
h_P \equiv \frac{h_{\sigma}\,h_r}{2h_r + h_{\sigma}}.
\end{equation}

\noindent Empirically, the gas density falls far more steeply with
radius than the velocity dispersion does
(Section~\ref{gasdenveldispsec}).  Therefore, the $r/h_r$ term will
typically dominate, such that $h_P\approx h_r$.

Alternatively, we can attempt to estimate the pressure directly from the gas
surface density.  \citet{joung2009} use hydrodynamic grid simulations to
calculate the turbulent pressure in stratified gas layers, after
adopting a power-law relationship between gas density and star
formation rate.  They find that the turbulent pressure scales with the
star formation rate per unit area $\Sigma_{sfr}$ like

\begin{equation}  \label{joungpressuresfreqn}
P_{turb}\propto \Sigma_{sfr}^{0.66\pm0.04},
\end{equation}

\noindent which, for a Schmidt law
\citep[$\Sigma_{sfr}\propto\Sigma^{1.4}$;][]{kennicutt1998} gives a
power-law relationship between the turbulent pressure and the local
gas surface density:

\begin{equation}  \label{joungpressureeqn}
P_{turb}\propto \Sigma^{0.92\pm0.05}.
\end{equation}

\noindent Note that the exponent differs from Joung et al.'s equation
20 for the relationship between turbulent pressure and the gas density
$\rho$, due to variations in the gas scale height with gas density and star
formation rate.  With equation~\ref{joungpressureeqn}, we convert
equation~\ref{vccorrpressureeqn} into a form where the strength of the
$\sigma^2_r$ term depends on gas density alone.

\begin{equation}  \label{vccorrjoungeqn}
V_c^2(r) = V_\theta^2(r) - \sigma^2_r(r) \, 
                  \left[\frac{0.92\,r}{\Sigma(r)}\,\frac{d\Sigma(r)}{dr}\right].
\end{equation}

\noindent For an exponential gas distribution, this equation reduces to

\begin{equation} 
V_c^2(r) = V_\theta^2(r) - \sigma^2_r(r) \, 
                  \left[\frac{0.92\,r}{h_r}\right],
\end{equation}

\noindent which is comparable to equation~\ref{expeqn} if $h_P \approx
h_r/0.92$.  For comparison, our first derivation of the correction to
$V_c$ (equation~\ref{hPeqn}) suggests that $h_P\approx h_r/1.4$ for
$h_{\sigma}\approx 5h_r$ (Section~\ref{gasdenveldispsec} below).  The
\citet{joung2009} prescription should therefore produce changes in
$V_\theta$ that are smaller than those in equation~\ref{expeqn} by
$\sim$60\%, for the same velocity dispersion and gas surface density
profile.  Note however that the \citet{joung2009} prescription in
equation~\ref{joungpressuresfreqn} has not yet been calibrated in the
low star formation efficiency regime, within which the exponent on
the relationship between gas surface density and star formation rate
is typically much steeper than that adopted for a Schmidt law
\citep[e.g.,][]{bigiel2008}.  

Unless otherwise stated, however, we will retain the more general form
of equation~\ref{vccorreqn} when calculating the effects of pressure
support.

\section{Corrections to the Mass Density} \label{massdenscalcsec}

The reduction of the tangential velocity has a direct impact on the
inferred mass density $\rho_{inferred}(r)$.  For an arbitrary rotation
curve, the density at some radius $r$ can be calculated from the rotation
curve and its derivative

\begin{equation} \label{rhoeqn}
\rho(r)=\frac{1}{4\pi G} \, \left(\frac{V_c}{r}\right)^2 \,
          \left(1 + 2\,\frac{{\rm d}\ln{V_c}}{{\rm d}\ln{r}}\right),
\end{equation}

\noindent assuming the potential is roughly spherical.

The mean mass density within a radius $r$, $\langle \rho(<r) \rangle$,
can also be derived from the rotation curve

\begin{equation} \label{meandenseqn}
\langle \rho(<r) \rangle = \frac{3}{4\pi G}\left(\frac{V_c(r)}{r}\right)^2.
\end{equation}

\noindent This quanity has the benefit of being a more robust
measurement of the density, since it depends only on the normalization
of the rotation curve, and not its derivative.

If no corrections are made for pressure support, then the 
mean density inferred to be within each radius is reduced from the
true density, since $V_\theta(r)<V_c(r)$.  For the
specific case of exponential scale lengths for the gas and velocity
dispersion,

\begin{equation} \label{meandenscorreqn}
\langle \rho_{inferred}(<r) \rangle = \langle \rho_{true}(<r) \rangle \,
               \left( 1  - \left(\frac{\sigma^2_r(r)}{v^2_c(r)}\right) \, 
                  \left[\frac{r}{h_P}\right]\right),
\end{equation}

\noindent confirming that the net result will be to infer a lower mean density than
the true mean density $\langle \rho_{true}(<r) \rangle$.  

Pressure support will also alter the apparent slope of the density
profile.  The power law slope of the density $\alpha$ (i.e.,
$\rho\propto r^\alpha$ such that $\alpha = \frac{{\rm
    d}\ln{\rho}}{{\rm d}\ln{r}}$) can be related to the logarithmic
slope of the rotation curve $\gamma$ (i.e., $\gamma = \frac{{\rm
    d}\ln{V_c}}{{\rm d}\ln{r}}$) through

\begin{equation}  \label{alphaeqn}
  \alpha = 2\,(\gamma-1) + 
           \frac{1}{1+\gamma}\,\frac{{\rm d}\gamma}{{\rm d}\ln{r}}
\end{equation}

\noindent Pressure support will tend to make the apparent rotation
curve shallower than the true rotation curve, reducing the value of
$\gamma$.  We therefore expect that neglecting pressure support would
lead one to infer shallower inner density profiles (i.e., smaller
values of $\alpha$), for realistic values of $\gamma(r)$.

The reduction in density will be most significant where the circular
velocity is small compared to the gas velocity dispersion, and will
thus be noticible over a larger fraction of a galaxy's radii when the
virial velocity of the galaxy is small.  Thus, pressure support is
potentially most significant in the regimes where the apparent
discrepancies from simulated dark matter halos are largest -- namely,
the inner regions of low mass galaxies.

\subsection{The Amplitude of Pressure Support}  \label{plotsec}

To calculate the amplitude of the effects of pressure support, we now
calculate changes to the rotation curve and the inferred density
profile.  We do so in two ways -- by starting with an assumed true
$V_c$ and then calculating the rotation curve $V_\theta$ that would be
observed, or by starting with observations of $V_\theta$ and then
inferring the true $V_c$.  We consider three baseline models, with
$v_{max}=30, 50,$ and $100\kms$.  For each of the models, we
first must set the physical parameters of the model galaxies
(i.e. scale lengths, velocity dispersions, etc).  In the following
sections, we discuss the observations used to constrain the scale
lengths and amplitudes of the gas density and velocity dispersion
profiles (Section~\ref{gasdenveldispsec}), the circular velocities of
realistic dark matter halos (Section~\ref{nfwsec}), and the observed
galaxy rotation curves (Section~\ref{arctansec}).

\subsubsection{Setting the Gas Density and Velocity Dispersion Profile}  \label{gasdenveldispsec}

To assign realistic values for the amplitude of the pressure gradient,
we must first specify the radial behavior of the gas density and the
velocity disperion profile (e.g., Equation~\ref{pressuredefneqn}).  We
can constrain these values using the wealth of H{\sc i} observations
of nearby galaxies.

Based on the observations of \citet{swaters2002a}, we assume that the
gas density declines as an exponential with a scale length of $h_r$.
We choose values for $h_r$ using observations of the H{\sc i} scale
length $h_{HI}$ as a function of $V_c\equiv W_{20}/2$ given in
\citet{swaters2002a} for a large sample of dwarf irregular galaxies.
We find that galaxies follow $\log_{10}(h_{HI}) = 0.5(V_c/100\kms) -
0.1$ for $10\kms<V_c<100\kms$, after updating the distances in
\cite{swaters2002b} with more recent determinations derived from the
tip of the red giant branch, and eliminating galaxies with $D>10\Mpc$,
for which distances are more uncertain.  The scatter around the mean
relation is $\sigma_{h_{HI}}=0.12$.  We note that the H{\sc i} scale
lengths measured by \citet{swaters2002a} typically ignored the very
inner regions, where the H{\sc i} profile rolls over to a nearly
constant surface density of $6-10\msun\pc^{-2}$ for late-type disks.
However, we consider it likely that the saturation of the H{\sc i}
surface density is due to the conversion of H{\sc i} into H$_2$
\citep[see Figure~8 of][for example]{bigiel2008}, in which case the
gas density should in fact continue rising to the center.  We thus
assume that $h_r\approx h_{HI}$.

Our next step is to set the radial velocity dispersion profile of the
galaxies using \citet{tamburro2009}'s recent analysis of H{\sc i} data
from THINGS \citep{walter2008}.  We take three galaxies (from their
sample of 11) as being representative of the different velocity
regimes we are probing -- Holmberg~II
\citep[$V_{max}\sim35$;][]{puche1992}, NGC~4214
\citep[$V_{max}\sim60\kms$;][]{allsopp1979}, and NGC~7793
\citep[$V_{max}\sim90\kms$;][]{carignan1990}; note that the exact
rotation speeds of these galaxies are uncertain, as
\citet{tamburro2009} chose nearly face-on galaxies for their velocity
dispersion analysis.  The velocity dispersion profiles are well fit
with exponentials $\sigma_{z}(r)=\sigma_{z,0}\exp{(-r/h_\sigma)}$, with
$\sigma_{z}$ measured approximately perpendicular to the plane of the
galaxy.  For our three representative galaxies, the velocity
dispersion profile of the gas is well fit with
$\sigma_{z,0}=17.2,18.2,$ and $18.2\kms$ and $h_\sigma=7.3,6.0,$ and
$11.5\kpc$, for HoII, NGC~4214, and NGC~7793, respectively.  There is
remarkably little variation in the central velocity dispersion
inferred for the range of galaxy masses we considered, and only a
modest variation of $h_\sigma$.  The scale lengths of the velocity
dispersion profiles are much longer than those inferred for the gas
surface density.  Using the relationship between rotation speed and HI
scale length we derived above, we estimate $h_\sigma/h_r\sim 4-6$.
Going forward, we therefore adopt $\sigma_{r,0}=18\kms$ for all model
galaxies, and $h_\sigma/h_r\sim5$.  Note that we are forced to
assume that the velocity dispersion is isotropic
(i.e. $\sigma_z=\sigma_r$), given that we lack any reliable way to
measure $\sigma_r$ for the gas.

\subsubsection{Pressure Support for a True NFW Circular Velocity} \label{nfwsec}

With the above profiles for the gas density and velocity dispersion,
we have sufficient information to calculate the radial pressure
gradient for our model galaxies.  We now must make assumptions for the
rotational velocities of the model galaxies.  In this section, we assume
that there is some true underlying circular velocity profile known from N-body
simulations, and then calculate the apparent rotation curve, including the
effects of pressure.  

We adopt a true $V_c(r)$ for the model galaxies by assuming that the
mass distribution is dominated by an underlying NFW density profile
\citep{navarro1996} with a virial concentration parameter $c_{vir}$.
We adopt virial velocities $V_{vir}$ that produce the desired fiducial
rotation speeds at $\sim\!10h_r$ (such that typically $V_{max}\sim1.25
V_{vir}$).  We then use the numerical simulation results of
\citet{maccio2008} to fix $c_{vir}$ as a function of $V_{vir}$, adopting
$\log_{10}(c_{vir}) = 1.579 - 0.283\,\log_{10}(V_{vir}/\kms)$ for the WMAP5
cosmology; for the three fiducial rotation speeds of 30, 50, \&
$100\kms$, we derive $c_{vir}\!=\!15.7$, 13.5, \& 11.0, respectively.  The
choice of $V_{vir}$ and $c_{vir}$ then fixes the core radius $r_c$ for the density
profile, where $r_c=r_{vir}/c_{vir}$ and
$(V_{vir}/75\kms)=(r_{vir}/151\kpc)$ for an assumed WMAP5 cosmology.
We then calculate the true circular velocity $V_c(r)$
expected for the NFW profile, and derive the apparent angular velocity
$V_\theta(r)$ using equation \ref{vccorreqn} to include the
effects of pressure support.

We plot the resulting observed rotation curves in
Figure~\ref{vcmodelnfwfig}, for the three fiducial rotation speeds.  At
high rotation speeds $V_c\gg\sigma_r$, the angular momentum support is
far more important than pressure support, and the observed rotation
curve is an excellent tracer of the true circular velocity.  At low
rotation speeds, however, pressure support becomes significant, such
that the gas disk can rotate more slowly while still
being supported against collapse.  In such cases, the observed
rotation speed is significantly less than the true circular velocity.

We calculate the density profiles that would be inferred from the
observed rotation curves.  We plot both the true and the inferred
density profiles in Figure~\ref{rhomodelnfwfig}, for the same rotation
curves shown in Figure~\ref{vcmodelnfwfig}.  As expected, the inferred
density profile is a good match to the true density profile for the
models where pressure contributes little to the overall support of the
disk.  At lower rotation speeds, however, pressure support can lead to
a significant reduction in the inferred density, by a factor of 1.46
in the $V_{max}=50\kms$ model, and 2.18 in the $V_{max}=30\kms$ model.
If we consider the $1\sigma$ scatter in the halo concentration from
\citet{maccio2008}, these reductions increase to factors of 1.96 and
6.98, respectively, at lower concentrations, producing more slowly
rising circular velocities, and thus larger radial ranges over which
the ratio of $\sigma/V_c$ is large and pressure support is
significant.  At $+\,1\sigma$ higher concentrations, the inferred
density is reduced by smaller factors of 1.25 and 1.50 in the
$V_{max}=50\kms$ and $30\kms$ models, respectively.  The change in the
inferred density is far less sensitive to our choice of
$h_\sigma/h_r$.

Figure~\ref{rhomodelnfwfig} suggests that pressure support can
potentially lead to inferred densities that are significantly lower
than the true underlying density, particularly for galaxies where the
velocity dispersion is a significant fraction of the rotation speed.
However, for the particular form of our model galaxies, there is no
significant change in the inferred inner slope of the density profile.
Both the true and the inferred density profiles show a cusped inner
density profile with a power-law slope of approximately -1.  Thus,
pressure support does not necessarily create the appearance of a cored
constant density mass profile, at least for our particular choice of
velocity dispersion profile.  However, while the literature suggests a
possible discrepancy between observations of the inner density profile
slope ($d\ln{\rho}/d\ln{r}$) and N-body simulations
\citep[e.g.,][]{deblok2009}, we note that such measurements depend on
the second-derivative of the observed rotation curve (e.g.,
equation~\ref{rhoeqn}), and are thus highly dependent on non-circular
motions \citep[e.g.,][]{spekkens2007,wada2002}, resolution, and
changes in inclination.

\subsubsection{Pressure Support for an Observed Isothermal Circular Velocity} \label{arctansec}

In the above example, we considered how pressure support could alter
the rotation curve and density profile expected for an NFW profile.
In this example, we consider the true circular velocity that would be
inferred for a frequently used parameterization of the
{\emph{observed}} rotation curve shape, after accounting for turbulent
pressure support.  In particular, we consider the arctangent
description of the observed rotation curve

\begin{equation}  \label{arctaneqn}
V_\theta(r) = V_t \, \left(\frac{2}{\pi}\right) \, {\rm{arctan}}(r/r_t), 
\end{equation}

\noindent following the notation in \citet{courteau1997}.  This parameterization
produces a rotation curve with an approximately linear rise out to a
radius $r_t$, beyond which it rolls over to a constant ``flat''
rotation curve with amplitude $V_t$.  The associated density profile
has a constant density core with $\rho_0\approx(3/\pi^3
G)\,(V_t/r_t)^2$, which then declines as
$\rho\propto r^{-2}$ beyond $r_t$.


We fixed values of $r_t$ as a function of $V_{\theta,max}$ using data
from \citet{courteau1997} (including the data in Tables 6 and 7, for
both the \citet{courteau1997} and \citet{mathewson1992} samples of
late type spirals), and from the \citet{spekkens2005} sample of
dwarfs.  The \citet{spekkens2005} rotation curves have been fit with a
slightly different functional form, but one that agrees extremely well
with equation~\ref{arctaneqn} if their distance scale parameters are
divided by 1.2 (i.e.\ $r_t\approx r_{ex}/1.2$ in the
\citet{spekkens2005} notation) and their velocity scale parameters are
divided by 0.87 ($V_t\approx V_{ex}/0.87$).  We approximate
$V_{\theta,max}$ using half of the H{\sc i} full-line velocity width
$W_{20}$ for all samples.  We use only those galaxies with
asymptotically flat rotation curves ($\beta=0$ for the
\citet{courteau1997} fits and $\beta<0.03$ for the
\citet{spekkens2005} fits, where $\beta$ is a free parameter that
accomodates rotation curves that do not become asymptotically flat at
large radii); restricting the sample to this subset allows the
\citet{spekkens2005} and \citet{courteau1997} samples to be easily
merged, in spite of their different parameterizations of the rotation
curve shape.

Qualitatively, $r_t$ increases for lower rotation speeds over the
interval from $120 < V_{\theta,max} < 300\kms$, as the effect of the
central bulge on the density profile becomes systematically less
important at lower galaxy masses.  Below $120\kms$, however, all
galaxies are essentially bulgeless, and the trend reverses, such that
$r_t$ becomes slightly smaller with decreasing rotation speed. Because
the bulge is essentially gone, the scaling of decreasing halo$+$disk
scale length with decreasing halo mass dominates at low masses.  Since
we are only interested in low mass galaxies with rotation speeds less
than $100\kms$, we fit only to this latter regime, and adopt
$\log_{10}(r_t/\kpc) = (0.0 \pm 0.25) + 0.3 \times
(V_{\theta,max}/100\kms)$ as a good representation of the range of the
data.  We note, however, that the scatter in this regime is quite
large.  Thus, at fixed maximum baryonic rotation speed (i.e.,
$V_{max}\approx W_{20}/2$), there is a significant range of density
concentrations, such that in some galaxies the rotation curve reaches
its peak value quickly, while in others the rotation curve is still
rising at its last measured point.

In Figure~\ref{vcmodelisofig} we show the fiducial observed arctan
rotation curves ($V_\theta(r)$) adopted for the three mass ranges,
along with the adopted velocity dispersion profiles (discussed in
Section~\ref{gasdenveldispsec} above).  We also show the true rotation
curves ($V_c(r)$) that would produce the observed rotation curve,
given an outward pressure force due to gas pressure (e.g.,
equation~\ref{vccorreqn}).  As we saw in Figure~\ref{vcmodelnfwfig}
for the NFW density profile circular velocities, the difference
between the true circular velocity and the observed rotation speed are
proportionally larger in lower mass galaxies.  In addition, the
differences between $V_\theta(r)$ and $V_c(r)$ are similar for a fixed
maximum rotation speed, whether the models began with a true NFW circular
velocity $V_c(r)$ (Figure~\ref{vcmodelnfwfig}), or with an observed arctangent
rotation curve $V_\theta(r)$ (Figure~\ref{vcmodelisofig}).  

Significant differences between the two approaches are seen, however,
when we examine the inferred density profiles (equation~\ref{rhoeqn})
for the artangent rotation curves, shown in
Figure~\ref{rhomodelisofig}.  The density profile derived from the
observed rotation curve would naively be interpreted as a modified
isothermal density profile with a constant density core.  However,
once the effects of pressure support are included, the true density
profile is seen to have a strong central cusp.  The differences
between the true and the observed density profiles are dramatic, and
as such, failure to take pressure support into account would lead to
fundamentally different scientific conclusions.

The relationship between the true and observed density profiles are
quite different in the two cases we considered.  When we started with
a highly cusped density profile and then calculated the observed
rotation curve in the presence of pressure support, we found little
difference between the true NFW density profile and that which would
have been inferred in the absence of corrections for pressure support.
Both showed steep inner power law cusps with
comparable slopes.  In contrast, when we started with an observed
arctangent rotation curve, the true density profile that would be
derived after correcting for pressure support is far more cusped than
one would infer from the observed rotation curve using traditional analyses.

Taken together, the two classes of models suggest that the typical
effect of correcting for pressure support will be to make the inner
density profile steeper, although by a small amount for apparent
rotation curves that already suggest a steep inner cusp.  More
notably, however, the inferred density profile can be incredibly
sensitive to the details of the pressure support.  The differences
between the observed $V_\theta$ rotation curves in
Figures~\ref{vcmodelnfwfig}~\&~\ref{vcmodelisofig} are rather subtle,
but they yield very different amplitudes for the correction needed to
derive the true density profile.  The corrections depend on the ratio
of $V_\theta$ to $\sigma_r$, and thus we expect an equal level of
sensitivity to subtle changes in the velocity dispersion profile.


This sensitivity is likewise reflected in the varying importance which
past authors have placed on pressure support.  Slight changes in
assumptions for the velocity dispersion profile (e.g., flat, gaussian,
exponential) led to a wide range of conclusions about the importance
of correcting for pressure.  Even more worrisome, this sensitivity
indicates the difficulty in ever making an accurate correction for
pressure support.  Minor deviations in $V_\theta$ due to non-circular
motions and varying inclination will easily compromise reconstruction
of the true density profile, although numerical experiments seem to
indicate that such affects are more likely to lead to overestimates of
the mass, rather than the underestimates produced by neglecting
pressure support \citep{wada2002}.  On the other hand, any correction is
likely to push the inferred density profile closer to the true density
profile.  We discuss further complications due to observational limitations
below in Section~\ref{obslimsec}.

\subsection{Reductions in the Mean Density} \label{alamsec}

For a spherical mass density, the mean interior density within some
radius $r$ is proportional to $(V_c(r)/r)^2$.  Thus, the mean density
inferred from a rotation curve will be artificially reduced if the
observed rotational velocity is less than the true circular velocity,
as it is in the presence of pressure support.  This effect can be seen
for both sets of models in
Figures~\ref{rhomodelnfwfig}~\&~\ref{rhomodelisofig}, which show
systematic reductions in the density that would be inferred if
pressure support were not taken into account.

The mean interior density of a galaxy has been recently characterized
by a dimensionless density $\Delta_{V/2}$, defined at the radius
$r_{V/2}$ where the rotation curve has risen to half its maximum
value $V_{max}$ (i.e., where $V_\theta(r_{V/2})=V_{max}/2$).  The dimensionless
density is defined at the mean interior density at this radius divided
by the current critical density of the Universe
\citep{alam2002}:

\begin{eqnarray}
\Delta_{V/2} &\equiv& \frac{\langle\rho(<r_{V/2})\rangle}{\rho_{crit}} \\
&=& 50 \, \left(\frac{V_{max}}{\kms}\right)^2 
       \, \left(\frac{h^{-1}\kpc}{r_{v/2}}\right)^2.
\end{eqnarray}

This quantity can be easily calculated for observed rotation curves
and for simulated dark matter halos.  Observationally, it is far more
robust to uncertainties than calculations of the inner slope of the
density profile, since it depends only on a single point on the
rotation curve, at radii where the rotational speed is typically much
larger than non-circular motions.

Initial comparisons between the value of $\Delta_{V/2}$ for simulated
halos and for low surface brightness galaxies suggested that the mean
interior density of the simulated CDM halos was much higher than
observed in real galaxies \citep{alam2002,zentner2003}, even when no
correction for the baryonic mass of the galaxies was made.  Recent
comparisons \citep{maccio2008} show less tension between the
predictions and observations, largely due to the reduction in the
predicted halo density with the currently favored WMAP5 parameters.
However, the two barred galaxies modeled in detail by
\citet{weiner2001} and \citet{zanmarsanchez2008} find values of
$\Delta_{V/2}$ that are still an order of magnitude below the WMAP5
CDM prediction (although some of the low halo density may be due to
angular momentum exchange from the bar itself; see review by
\citet{sellwood2009}).

In Figure~\ref{alamfig} we show the predicted measurements of the
dimensionless density $\Delta_{V/2}$ for our fiducial galaxies,
showing both the true value of the dimensionless density (solid
symbols), and the artificially low value that would be measured in the
presence of pressure support (open symbols).  For both the true NFW
models (circles) and the observed arctan rotation curve models
(triangles), pressure support significantly reduces the apparent
dimensionless density for low mass galaxies.  It has little effect for
$100\kms$ halos, however.  Thus, correcting for pressure support could
reduce any observed discrepancy for low mass halos, but would have
no signficant impact at higher masses.

\subsection{Can Pressure Support be Maintained?}  \label{shrinkingsec}

The pressure support discussed above depends on the surface density of
the gaseous disk.  However, star formation will tend to erode the gas
surface density, and thus reduce the pressure support.  The gas will
preferentially be depleted in the inner regions, due to the non-linear
dependence of the star formation rate on gas density.  The consumption
of gas will then lead to an inward flow of gas, which in turn will
help restore the pressure support.  In this section we evaluate the
importance of these effects, to assess whether pressure support can be
a long-lived phenomena in an evolving star forming disk.

The evolution of pressure support in a disk depends to first order on
the evolution of the gas surface density $\Sigma_g$.  The gas density
at a radius $R$ in a rotating disk is governed by a series of
differential equations governing the flow of matter through the
annuli.  The first of these equations describes the conservation of
matter during flows:

\begin{equation}   \label{massconservationeqn}
\frac{\partial \Sigma_g}{\partial t} + \frac{1}{R}\,\frac{\partial(R V_r \Sigma_g)}{\partial R}
 = -\frac{\partial \Sigma_\star}{\partial t},
\end{equation}

\noindent where the gas surface density changes in response to the
flow of matter between annuli (the second term on the LHS) and the
conversion of gas into stars, as measured by the change in the stellar
surface density $\Sigma_\star$.

The second equation describes the evolution of the radial forces,

\begin{equation}   \label{radialforceevolutioneqn}
\frac{\partial V_r}{\partial t} = \frac{V_\theta^2}{R} + \frac{\partial \Phi}{\partial R} - \frac{1}{\rho_g}\,\frac{\partial P}{\partial R},
\end{equation}

\noindent which allows small radial velocities to develop when the
angular velocity $V_\theta$ and pressure $P$ are no longer sufficient
to balance the force that results from the gravitational potential
$\Phi$.  We have neglected the radial advection of $V_r$, as terms
proportional to $V_r^2$ are small.

The third equation describes constraints from angular momentum conservation

\begin{equation}   \label{angularmomentumeqn}
\frac{\partial (R V_\theta \Sigma_g)}{\partial t} + \frac{1}{R} \, \frac{\partial (R V_r \cdot R V_\theta \Sigma_g) }{\partial R} = 
   - R\,V_\theta\,\frac{\partial \Sigma_\star}{\partial t},
\end{equation}

\noindent where the change in the angular momentum of the gas (first
term) is due to angular momentum flux through the annulus at a rate
controlled by the radial velocity $V_r$ (second term), and to the rate
at which angular momentum drops out of the gas phase and into the
stellar phase due to star formation (right hand side).  In this
equation we have ignored the effects of viscosity; see
\citet{firmani1996} for a full set of evolutionary equations
containing viscous terms.  Note that a gas disk will tend to collapse
inwards in the presence of star formation
(eqn.~\ref{radialforceevolutioneqn}), due to decreased pressure
support resulting from the conversion of gas into stars
(eqn.~\ref{massconservationeqn}).  However, the resulting inward flow
of high angular momentum gas (eqn.~\ref{angularmomentumeqn}) will
increase the circular speed, and thus will help counteract the
shrinking of the disk.

To solve this system of equations, we make a series of
simplifications.  We first assume that the star formation rate surface
density follows a Schmidt-type law, and is thus proportional to a
power $\alpha_{SF}$ of the gas surface density, such that ${\partial
  \Sigma_\star}/{\partial t} = \psi_{sfr}
(\Sigma_g/\Sigma_0)^{\alpha_{SF}}$ where $\Sigma_0$ is a fiducial
surface mass density and $\psi_{sfr}$ is the fiducial star formation
rate surface density when $\Sigma_g=\Sigma_0$.  We adopt
$\Sigma_0=10\Msun/\pc^2$, $\psi_{sfr}=4.07\Msun/\Gyr/\pc^2$, and
$\alpha_{SF}=1.4$, in agreement with the mean trends found by
\citet{bigiel2008}.

For our second assumptions, we assume that the gravitational potential
is unchanged by the evolution in $\Sigma_g$ and $\Sigma_\star$, since
the potential will be dominated by the dark matter and the stellar
disk.  We also assume that the gas disk starts in gravitational
equilibrium (such that ${\partial V_r}/{\partial t}=0$ initially)
before gas consumption drops the pressure support and takes the gas
out of equilibrium.

Third, we assume that because $h_{r}/h_\sigma\approx5$, radial
variations in the pressure are driven primarily by the change in gas
density with radius.  We thus assume that the velocity dispersion
varies slowly enough with radius that it can be approximated as a
constant, such that $({1}/{\rho_g})\,{\partial P}/{\partial R} \approx
({\sigma_r^2}/{\Sigma_g})\,{\partial \Sigma_g}/{\partial R}$.  

Finally, we assume that the velocity dispersion is constant with time,
and thus that the energy in the turbulence is continually replenished
due to energy input from evolving stars and magneto-rotational
instabilities \citep[see reviews by][]{maclow2004,elmegreen2004}.  To
first order, this assumption has empirical support from observations
that the average velocity dispersion depends very weakly on star
formation rate, varying by less than a factor of two when the average
star formation rate per unit area varies by more than a factor of 1000
\citep[c.f., Figure~2 of ][]{dib2006}, for star formation rates
comparable to those seen in late type galaxies.  However, the
assumption of constant velocity dispersion may be less valid in
regions with star formation rates higher than $>10^{-3}\,{\rm
  M_\odot\,yr^{-1}\,kpc^{-2}}$, for which the apparent velocity
dispersion does appear to correlate with star formation rate
\citep[c.f., Figure~1 of ][]{tamburro2009}, although again with only a
factor of two range in velocity dispersion.  Thus, although we expect
that the decline in star formation rate due to gas consumption could
potentially reduce the turbulent velocity dispersion, and thus the
degree of pressure support, we expect the temporal variation in
$\sigma$ to be modest, due to the very weak empirical correlations
between velocity dispersion and all other physical quantities.

With the above assumptions, we can rewrite the system of equations as:

\begin{eqnarray}
\frac{1}{\Sigma_g}\,\frac{\partial \Sigma_g}{\partial t} &=& 
   - \frac{\psi_{sfr}}{\Sigma_0}\,\left(\frac{\Sigma_g}{\Sigma_0}\right)^{\alpha_{SF} -1} - \frac{V_r}{R}\left[ 1 + \frac{R}{V_r}\,\frac{\partial V_r}{\partial R} + \frac{R}{\Sigma_g}\,\frac{\partial \Sigma_g}{\partial R} \right] \\
\frac{1}{V_\theta}\,\frac{\partial V_\theta}{\partial t} &=& 
   - \frac{V_r}{R}\left[1 + \frac{R}{V_\theta}\,\frac{\partial V_\theta}{\partial R} \right]\\
\frac{\partial V_r}{\partial t} &=& 
   a_{g,0} + \frac{V_\theta^2}{R}\left[1 - \left(\frac{\sigma_r}{V_\theta}\right)^2\frac{R}{\Sigma_g}\,\frac{\partial \Sigma_g}{\partial R}\right]
\end{eqnarray}

\noindent where $a_{g,0}$ is the initial gravitational acceleration,
set to keep the system in balance at the initial timestep (i.e. such
that ${\partial V_r}/{\partial t}\!=\!0$ at $t\!=\!0$).  These form a
system of three differential equations with three unknowns
($\Sigma_g(t)$, $V_\theta(t)$, and $V_r(t)$), which can be integrated
forward from some assumed initial condition.  We adopt an exponential
profile for $\Sigma_g(t\!=\!0)$, with scale length $h_r$ chosen
according to $V_{max}$, as described for our fiducial disks in
\S\ref{gasdenveldispsec}.  We parameterize the initial rotation curve
with an arctangent (eq.\ \ref{arctaneqn}), with a scale length chosen
as described in \S\ref{arctansec}.  We assume that the disk is
initially in rotational equilibrium ($V_r(t\!=\!0)\!=\!0$), and fix
the radial gas velocity dispersion at $\sigma_r\!=\!15\kms$.  Although
it is unlikely that a real galaxy would ever be found in such an
idealized state, these initial conditions are adequate for determining
the timescales over which pressure support is significant.

In Figure~\ref{diskevfig} we show the resulting evolution of the
disk's gas surface density (left column) and rotation curve (right
column), for fiducial rotation speeds of
$V_{max}=30$,\,50,\,\&\,$100\kms$ (top to bottom), at a series of
timesteps separated by $1\Gyr$ (light to dark, with the darkest line
being the final state).  In all disks, the gas surface density
declines with time, with proportionally larger reductions in the
center due to the higher star formation efficiency at high gas densities.

The models also show noticable radial redistribution of gas in the
lowest mass galaxies, where pressure support is significant.  In these
galaxies, conversion of gas into stars leads to a reduction of pressure support.  
The reduced support then leads to an inward flow of
gas, causing the gas disk to evolve with a nearly self-similar surface
density distribution.  This inward flow of material raises the mean
angular momentum at each radius, causing the rotation speed to
increase with time.  In contrast, galaxies with higher rotation speeds
have lower degrees of pressure support, leading to little radial
redistribution of gas, and no obvious changes in the rotation curve.

Although the decline in the gas surface density does lead to some
radial shrinking of the outskirts of pressure-supported gas disks, the
rate of shrinking is not dramatic.  The radius at which the outer disk
passes through a fixed surface density (say,
$\Sigma_g\sim0.1\Msun/\pc^2$) falls by only 10\% over $5\Gyr$ for the
most pressure-supported galaxy, compared to 6\% for the least pressure
supported disk.  Thus, while the reduction in pressure support in
slowly rotating galaxies increases the rate of shrinking by a factor
of two compared to rapidly rotating galaxies, the absolute amount of
shrinking is small in either case.  In constrast, the shrinking of
disks due to falling pressure support was identified by
\citet{stinson2009} as a significant effect in the early evolution of
dwarf galaxies.  The rates calculated here are much smaller than found
in the simulations, however, presumedly due to some combination of
more rapid gas consumption and a larger degree of pressure support in
the simulations.

The shrinking of the disk helps to refuel gas to the central regions.
However, like the shrinking in the outskirts, the effect is small.
The final surface density in the $V_{max}=100\kms$ disk is
17\% of the initial value, but is somewhat higher (21\%) in the most
pressure supported disk ($V_{max}=30\kms$) due to the gas that has
flowed inwards as the pressure support has decreased.

The overall conclusion from Figure~\ref{diskevfig} is that pressure
support can have an impact on the evolution of gaseous disks, but the
overall size of the impact should be modest.  Moreover, while the degree of
pressure support does decline somewhat with time, the overall decline
is small, as can be seen from the fact that the ratio of
$\sigma_r/V_\theta$ remains close to its initial value.  Thus, a disk
that is partially supported by turbulent pressure is likely to be so
over its lifetime, barring any significant change in its gas
distribution or star formation rate in response to external forces
(tidal interactions, infall, etc).

One caveat to the above conclusion is that the turbulent velocity of
the gas may be time-dependent, as opposed to constant as we assumed
above.  For example, if the turbulent velocity of the gas is
proportional to the star formation rate \citep[e.g.,][]{tamburro2009},
then pressure support should increase during times of high star
formation activity, and decrease during quiescent intervals.  Such a
variation could produce epochs of disk expansion and contraction,
correlated with periods of high and low star formation rates,
respectively.  Empirically, however, the turbulent velocity does not
seem to vary by more than a factor of two from galaxy-to-galaxy, even
when the star formation rate per unit area varies by factors of ten or
more.  Self-regulation may therefore limit the strength of any
possible correlation between turbulent velocity and star formation
rate, which would then keep the degree of pressure support roughly
constant.

\section{Application to Real Galaxies}  \label{datasec}

The calculations above adopt idealized model galaxies.  We now
derive the amount of pressure versus rotational support for real
galaxies, using publicly-available H{\sc i} data.

Estimating the amplitude of pressure support requires measurements of
the rotation speed, turbulent velocity, and gas surface density.
Unfortunately, these quantities are frequently difficult to constrain observationally,
especially within a single galaxy.  First, the rotation speed is best
measured for highly inclined galaxies with well-determined
inclinations, but the velocity dispersion and gas surface density are
most reliably measured for face-on systems.  Second, the radial
velocity dispersion controls radial pressure support, but only the
vertical velocity dispersion can be easily measured (although the
radial component can potentially be derived from detailed modelling).
Third, the balance between pressure and rotation support can be most
easily evaluated for velocity fields with negligible non-circular
motions and constant inclination with radius, but such systems are
rare among low mass galaxies.  Third, the measured gas surface density
should include both the molecular and atomic components, but molecular
gas is rarely detected in low mass galaxies, possibly due to a strong
mass-dependent variation in the ratio of CO to H$_2$
\citep[e.g.,][]{boselli2002}.

We select galaxies for testing the amplitude of pressure support from
the THINGS sample of high-resolution VLA observations of H{\sc i} \citep[The
H{\sc i} Nearby Galaxy Survey;][]{walter2008}.  The data cubes for THINGS
galaxies are publicly available, and rotation curves \citep{deblok2008}
and velocity dispersion profiles \citep{tamburro2009} are available for
the majority of the galaxies.  We restrict our initial selection to
galaxies with $W_{20}\lesssim 100\kms$, since only galaxies with
$V_{max}\lesssim 50\kms$ are likely to show significant pressure
support.  The candidates are M81dwA, HoI, DDO53, M81dwB, HoII, NGC
628, NGC 4214, DDO154, and NGC 2366, in order of increasing apparent
rotation speed (uncorrected for inclination).
We further restrict the sample to those with inclinations $>40^\circ$,
for which rotation curves can be reliably measured; only these
galaxies have had rotation curves published in \citet{deblok2008}.  The
remaining galaxies are DDO~154 ($V_{max}\sim50\kms$) and NGC~2366
($V_{max}\sim60\kms$), both of which are in a regime where the effects
of pressure support may be marginally detectable.  However, of these,
NGC~2366 has a strongly declining rotation curve, suggestive of a
kinematically disturbed galaxy, leaving DDO~154 as the only candidate
for providing a legitimate analog of the models discussed in this paper.
The few galaxies with potentially lower rotation speeds are nearly
face-on, have asymmetric velocity fields, or are nearly spatially
unresolved.

The rotation curve of DDO~154 was kindly given to us in
electronic format by Erwin de Blok, based on \citet{deblok2008}.  We
maintained consistency with the \citet{deblok2008} rotation curve by
adopting their values for dynamical centers, inclinations, and
distances for the sample galaxies, using the natural-weighted data
sets. 

We estimate the average H{\sc i} surface density $\Sigma_{HI}$ and
velocity dispersion $\sigma_{HI}$ as functions of galactic radius.  We
obtained the inclination-corrected surface density from the moment~0
map, and the velocity dispersion from the moment~2 map, by integrating
elliptical annuli with a width equal to that used by
\citet{deblok2008} to derive the rotation curve.  The total gas
surface density at each radius was assumed to be
$\Sigma_{gas}(r)=1.4\times\Sigma_{HI}(r)$, where the factor of 1.4 is
an approximate correction for metals and molecular gas (since the
latter is currently unavailable for these galaxies).  We also
considered a more extreme correction for molecular gas by assuming
that $\Sigma_{gas}$ follows an exponential profile into the center of
the galaxy, such that the outer regions are dominated by atomic gas,
but the inner regions are dominated by molecular gas, producing the
apparent roll-over of $\Sigma_{HI}$ towards the center.  Both of the
resulting gas density profiles are shown in Figure~\ref{ddo154fig}.

The pressure at each radius was taken to be
$P(r)=\langle\sigma_{HI}^2(r)\Sigma_{gas}(r)\rangle$, which implicitly assumes that
$\sigma_r \approx \sigma_{HI}$.  This assumption is not necessarily
true, given that interaction between the ISM and stellar feedback may
well differ between the vertical direction (where the ISM becomes
diffuse over $100\pc$ scales) and the radial and tangential directions
(where the ISM remains dense on kiloparsec scales).  This calculation
also assumes that the pressure is due to random motions of the gas,
rather than locally coherent motions due to expanding superbubbles.
To better capture the broad radial trend in the pressure gradient,
we chose to smooth the velocity dispersion and surface density before
calculating the pressure.  We then use equation~\ref{vccorreqn} to
calculate the true circular velocity that would be observed in the
absence of pressure support.

In Figure~\ref{ddo154fig} we plot the resulting data for DDO~154.  The
right hand panel shows the velocity dispersion profile (dotted line),
the observed tangential velocity (black solid line), and the inferred
underlying circular velocity (red), both for the observed gas surface
density (solid line) and for an exponential surface density
extrapolated to the center (dashed lines).  There is little evidence
that pressure support plays any significant role in this particular
galaxy.  The pressure gradient appears to be negligible in the inner
regions, due to the relatively flat velocity dispersion profile and
inner gas surface density profile.  Moreover, the H{\sc i} velocity
dispersion has a low amplitude compared to the fiducial model
($\sim10\kms$ versus $18\kms$).  There is some evidence for pressure
support in the outer regions, for which the true rotation speed may be
$\sim$10\% higher than observed.  However, we have verified that this
difference makes no significant changes in the inferred density
profile.  Likewise, the inferred dimensionless density $\Delta_{V/2}$
increases by less than 2\%, to ${\rm log}_{10}\Delta_{V/2}=5.26$, which
still lies below the fiducial curve from simulations in
Figure~\ref{alamfig}.  However, while the evidence for pressure
support is small in this case, this galaxy is not in a regime where we
expected pressure support to be obvious.

\section{Discussion of Possible Limitations}   \label{discussionsec}

Our analysis of pressure support reveals a number of contradictions.
On the one hand, straight-forward analytic calculations demonstrate
that pressure support is likely to contribute significantly to the
radial support of low mass galaxy disks, in agreement with numerical
simulations.  On the other hand, observational evidence for pressure
support is weak.  Our analysis of DDO~154 finds that corrections for
pressure support would make only minor changes to the interpretation
of the observed rotation curve.  \citet{swaters2009} reaches a similar
conclusion for the dwarf irregulars in the \citet{swaters2002a}
sample, which have $V_c\!\gtrsim\!60\kms$.

These apparent contradictions must arise from limitations in the
various analyses.  Each of the approaches for assessing pressure
support come with a number of caveats, which we now discuss.

\subsection{Limitations of Observations of Pressure Support} \label{obslimsec}

Assessing the importance of turbulent pressure support requires
knowledge of the radial turbulent velocity dispersion, the gas surface
density, and the tangential rotation speed.  Unfortunately,
observations cannot measure all of these quantities within a single
galaxy.  Instead, we are forced to make several approximations, none
of which may be completely satisfactory.  

The first approximation is the use of the line-of-sight H{\sc i} velocity
dispersion as a substitute for the radial velocity dispersion.  There
is no reason to believe that the velocity dispersion of the gas is
strictly isotropic, given the very different boundary conditions for
motions within the plane compared to motions in the vertical
direction.  The amplitude of $\sigma_r$ may therefore be
systematically different from the line-of-sight dispersion measured in
nearly face-on galaxies.  Adaptive mesh simulations of rotating disks
by \citet{agertz2009} find that the velocity dispersion in the plane
of the galaxy is at least a factor of two higher than the vertical
component of the velocity dispersion.  Because the importance of
pressure support scales as $\sigma_r/V_c$, using the measured vertical
velocity dispersion ($\approx\!\sigma_z$) may significantly
underestimate the degree of pressure support, as can be seen from the
prefactor of the term in brackets in equation~\ref{vccorrpressureeqn}.
Assuming a factor of two difference between $\sigma_r$ and $\sigma_z$,
the degree of pressure support for a galaxy with $V_c\!\sim\!70\kms$
may actually be as high as that calculated for a galaxy with
$V_c\!\sim\!35\kms$.  Note that the term within the brackets in
equation~\ref{vccorrpressureeqn} will be unlikely to change
significantly if $\sigma_z$ is used as an estimate for $\sigma_r$,
since (1) the logarithmic gradient in $\sigma$ is likely to be similar
between the two directions, assuming that the shape of the velocity
ellipsoid varies slowly with radius; and (2) the gradient in $\sigma$
contributes less to the pressure gradient than the gradient in surface
density, and thus uncertainties in the velocity dispersion gradient
are not likely to dominate the error budget.

Also concerning is that the line-of-sight H{\sc i} velocity dispersion only
tracks one particular phase of the gas.  Thus, H{\sc i} velocity dispersions
are likely to be biased tracers of the velocity dispersion in regions
of the galaxy that have significant molecular or ionized gas
fractions.  These other phases are known to have different scale
heights than the H{\sc i} \citep[see review by][]{vanderhulst1996}, and thus
must have different characteristic velocity dispersions.
Unfortunately, data on these other phases have historically been far
more difficult to obtain than HI.  Some hope for using only HI
observations comes from recent grid-based hydrodynamic simulations by
\citet{joung2009}, who find that different gas phases tend to come
into turbulent pressure equilibrium (mirroring what has long been
known for their thermal pressures).  Thus, measuring the atomic
turbulent pressure alone may provide an indirect estimate of the
pressure from molecular and ionized phases.

In addition to the challenge of measuring turbulent velocities, there
are also difficulties in comparing the observed tangential velocities
to analytic calculations, due to the presence of non-circular motions.
Analytic calculations assume that the gas moves on perfect circular
orbits, and thus has no radial streaming motions which could otherwise
support the radial extent of the disk.  However, most late-type
galaxies show some evidence for non-circular motions at the 5-$20\kms$
level \citep[e.g.,][]{gentile2005,trachernach2008}, which is
comparable to the typical amplitudes of turbulent velocities.  Thus,
when significant non-circular motions are present, the assessment of
the importance of pressure support becomes more complicated than
assumed in most calculations.  Unfortunately, it seems that
non-circular motions are almost always present in dwarf galaxies,
particularly at the low masses where pressure support is most likely
to be significant.  In our search for a suitable galaxy to
compare with our calculations (Section~\ref{datasec}), it proved to
be difficult to find low mass galaxies for which the velocity fields
do not appear to deviate from what is expected for a simple rotating
disk.  Signatures of warps and elliptical streaming motions were
sufficiently common that only one system barely qualified.

The difficulty of finding approrpiate systems for evaluating pressure
support is even worse in lower mass galaxies than DDO~154.  Although
these systems would potentially have the most dramatic evidence for
pressure support, they also have increasingly chaotic velocity fields
\citep[see ][]{lo1993,begum2008}, making the measurement of the
tangential velocities as complicated as measuring the radial velocity
dispersion.  Furthermore, the gas disks of low mass galaxies
frequently appear to be far from equilibrium.  The presence of
dramatic H{\sc i} holes \citep[e.g., see reviews
by][]{brinks2007,vanderhulst1996} and complicated velocity fields
indicates that the entire basis of the calculation (i.e., a disk in
equilibrium) may be suspect.

There are equally large limitations in measuring the gas surface
density.  The easiest gas component to measure is atomic Hydrogen
through the 21cm line.  However, molecular and ionized gas also
contribute to the turbulent pressure, but are far more difficult to
measure.  Molecular gas in particular is extremely challenging to
identify in late-type galaxies, for which there are few CO detections,
in spite of copious evidence for star formation
\citep{israel1995,matthews2005, das2006,leroy2005}.  The lack of CO
detections most likely arises from factor of ten variations in the
$X_{CO}$ conversion factor from CO to H$_2$, rather than any lack of
molecular gas \citep[e.g.,][]{ohta1993,arimoto1996,israel1997,
  madden1997,boselli2002,leroy2007}.  Thus, the true contribution of
H$_2$ to the turbulent pressure is particularly difficult to constrain
in the systems where it is most important.  Approximating the
molecular gas distribution by scaling the H{\sc i} observations is
unsatisfactory as well, since the ratio of H$_2$ to H{\sc i} is known to
vary spatially within galaxies \citep[e.g.,]{wong2002,leroy2008}.

The ionized gas component is as difficult to constrain as the
molecular.  Hot SN-heated gas is likely to fill the H{\sc i} ``holes''
that are prevalent in dwarf galaxies.  However, x-ray detections of
such gas are prohibitively expensive.  On the other hand, hot gas
appears to be a small fraction of the total mass in the Milky Way disk
\citep[e.g.,][]{ferriere1998}, suggesting that it may not necessarily
be a significant source of pressure in dwarfs, provided one can safely
extrapolate these results to such a different mass regime.

\subsection{Limitations of Simulations of Pressure Support} \label{simlimsec}

The pressure support in a stable galaxy disk depends on the properties of the
interstellar turbulence.  This turbulence
is thought to be driven by various forms of stellar feedback (winds,
SNe, etc), which inject energy into the ISM over a range of physical
scales \citep[see reviews by][]{elmegreen2004,maclow2004}.  This
energy then cascades to an even wide range of scales, forming a broad
power spectrum of turbulent energies.  The pressure that results from
the turbulence thus depends on the sum of kinetic energy across a wide
range of length scales.

Unfortunately for simulations, it appears that the vast majority of
the kinetic energy is due to tubulent motions on scales smaller than
$\sim$200$\pc$ \citep{joung2006}.  These small scales have been
successfully resolved using either adaptive mesh refinement (AMR) or
smoothed particle hydrodynamic (SPH) codes \citep[e.g.,][]{price2010},
but only for simulations that model a small portion of the disk.
While these high-resolution simulations can resolve the physical
scales that dominate the turbulent energy, the same codes have
difficulty simultaneoulsy simulating the large scale behavior of the
galactic disk and its cosmological environment.  In constrast,
simulations that capture the physics of the disk on larger scales
cannot currently fully resolve the ``subgrid physics'' of turbulence.

To explore the suitability of simulations for assessing the amplitude
of pressure support, we will consider two state of the art
simulations.  The first is the suite of SPH simulations used by
\citet{valenzuela2007} to argue for the significance of pressure
support, feedback, projection effects, and non-circular motions in
explaining the discrepancy between observed rotation curves and the
canonical NFW profile.  The second is the AMR simulation of
\citet{joung2009} used to assess the scaling between pressure and star
formation rate in self-gravitating stratified gas layers.

The \citet{valenzuela2007} SPH simulations modeled a moderately low
mass dwarf galaxy, created to be an analog of nearby dwarf irregular
galaxies ($V_c\sim 70\kms$).  Their initial conditions embedded a
mixed gas$+$star disk within a live axisymmetric NFW dark matter
halo.  Their model was then evolved using the SPH code GASOLINE
\citep{wadsley2004} for $1\Gyr$.  Because their simulation considered
an isolated disk evolved for a short period of time, the galaxy is
unaffected by any larger scale gravitational forces or long term
evolution driven by interactions with other galaxies.  These
restrictions allowed them to use a large number of particles
(2~million) and a small softening length ($30\pc$), such that
Newtonian forces were reached beyond $60\pc$.  The resulting gas disk
successfully reproduces the holes and filaments that are
characteristic of many dwarf galaxies, and has a slowly rising
rotation curve, in spite of being embedded within a true NFW profile.

Unfortunately, even these carefully crafted simulations do not seem to
be fully capturing the effects of pressure support.  The midplane
velocity field (their Figure~1) shows coherent rotational flow with
significant radial velocities in the inner regions. However, it shows
none of the fully turbulent motion that likely dominates the
interstellar pressure.  Comparing their effective force resolution
($\sim60\pc$) to the scales where the turbulent kinetic energy
dominates ($\lesssim200\pc$) suggests that they are unlikely to have
resolved the full amplitude of the turbulent pressure\footnote{Note
  that the difficulty in resolving the pressure is likely to become
  even more severe at higher star formation rates, which appear to
  shift the kinetic energy spectrum to even shorter characteristic
  length scales \citep{joung2009}.}.  Thus, their success in reproducing
slowly rising rotation curves is likely due to the dominance of the
other effects they identified, rather than turbulent pressure.  While this work
leaves open the question of the amplitude of pressure support, it does
point to mechanisms which may well cause the apparent discrepancies in
higher mass galaxies where pressure is expected to be negligible.

In constrast, the \citet{joung2009} AMR simulations do a superb job of
resolving the turbulent motions down to very small scales
($\sim$2$\pc$).  However, this high spatial resolution comes at the
expense of not modeling the full galactic disk.  Instead, they
simulate small patches of a vertically stratified gas layer, $0.5\kpc$
on a side, with periodic boundary conditions in the plane.  These
simulations therefore are unable to capture radial flows or shear
fields, and must view the disk in steady state.  There are also
several limitations in the simulation, such as the lack of correlation
between local density and the local SN rate, which may contribute to
the simulations' apparent failure to reproduce the large holes seen in HI
observations.  

On the other hand, the \citet{joung2009} simulations do offer a number
of lessons for thinking about turbulent pressure in general.  First,
both the thermal and turbulent pressures appear to be in pressure
equilibrium across phases, such that the turbulent pressure of the hot
gas is comparable to the turbulent pressure of the cold gas on the
larger scales where the turbulent pressure dominates.  The net result
is that pressure shows much lower variation with position than either
density or temperature.
Second, the turbulent pressure scales with the star formation rate,
since both are correlated with energy input from SNe.  This scaling
relationship provides a means to potentially capture the subgrid
physics of turbulence, which is otherwise unresolved in SPH
simulations \citep{joung2009}.  The next generation of SPH simulations
may thus have the power to capture both the large scale galactic
context provided by SPH, while better assessing the degree of pressure
support.

\subsection{Limitations of Analytic Calculations of Pressure Support} \label{calclimsec}

The analytic calculations presented in Section~\ref{pressuremathsec}
require a number of compromises that are made for analytic expediency,
but that probably fail to capture some of the relevant physics.
First, as discussed in Section~\ref{obslimsec}, gas in real low mass
galaxies shows significant deviations from pure circular motion, in
contrast to what is assumed in Section~\ref{pressuremathsec}.  Bulk
non-circular motions add an additional term to the Jeans equation,
which is not accounted for in our calculations.  The simulations in
\citet{valenzuela2007} show a significant contribution from
non-circular motion, that leads to a reduction in the inner rotation
curve. Second, the presence of holes and warps suggest that low mass
galaxy disks are frequently not in a steady-state equilibrium, again
in conflict with the analytic assumptions.  Third, the derivations in
Sections~\ref{pressuremathsec}~\&~\ref{massdenscalcsec} assume that
the gas disk has no significant viscosity, velocity anisotropy, or
tilt in the velocity ellipsoid.  Finally, we ignore any support due to
thermal pressure, which dominates only on smaller scales and lower
pressures 

\section{Conclusions}   \label{conclusionsec}

The calculations above suggest that in low mass galaxies with
$V_c\lesssim50\kms$, gas pressure may help to support the disk against
the inward force of gravity.  In such cases, the observed tangential
velocity of rotation curve tracers like H{\sc ii} or H{\sc i} will be
smaller than the circular velocity of an idealized test particle,
leading to underestimates of the mass contained within a given radius.
Accounting for pressure gradients in the gas may lead to higher mean
densities and steeper inner halo profile slopes for low mass galaxies,
thus reducing possible conflicts with $\Lambda$CMD on small scales.
Pressure support can be maintained over long time scales, even in the
presence of star formation.  In such cases, the reduction in pressure
support due to central star formation leads to an inward flow of gas,
which then re-establishes pressure support.

However, while our calculations suggest that pressure gradients could
help support the centers of low mass galaxies, we find that there are
many obstacles to ever establishing convincing observational or
numerical evidence for pressure's importance.  Pressure support
depends on the radial velocity dispersion of the entire ISM, but
measurements are typically limited to measuring only the vertical
velocity dispersion of the H{\sc i}.  In addition, the effects of
complex non-circular motions and evolving superbubbles may dominate
the structure of the gas disk, making corrections based on simple
steady-state circular models invalid in the low mass galaxies within
which we expect pressure to have the largest impact.  Likewise,
simulations currently do not have sufficient resolution to
simultaneously capture both the large scale galaxy dynamics and the
small scale interstellar turbulence which dominates the pressure.  In
light of these caveats, while we consider pressure support to be an
intriguing mechanism to reduce tensions with $\Lambda$CDM on small
scales, it is by no means a conclusive solution to current
concerns. Other proposed mechanisms \citep[e.g.,][]{governato2010} may
play an equally or more important role.

\acknowledgements 

The authors are very happy to acknowledge helpful discussions with
their colleagues Jeremiah Murphy, Mordecai Mac-Low, Tom Quinn, Rok
Roskar, and Fabio Governato.  They also warmly thank Erwin de Blok for
supplying several THINGS rotation curves in digital format.  The referee
is also thanked for suggestions that improved the paper.  J.J.D.\
acknowledges the hospitality of the Max-Plank Institute f\"ur
Astronomie in Heidelberg and Caff\'e Vita during part of this work.
A.S. and J.J.D. were partially supported by NSF grant 0807515.
This work made use of THINGS, ``The H{\sc i} Nearby Galaxy Survey''
\citep{walter2008}.

{\it Facility:} \facility{VLA}

\bibliographystyle{apj}  


\ifprintfig

\clearpage

\begin{figure}[t]
\centerline{
\includegraphics[width=2.25in]{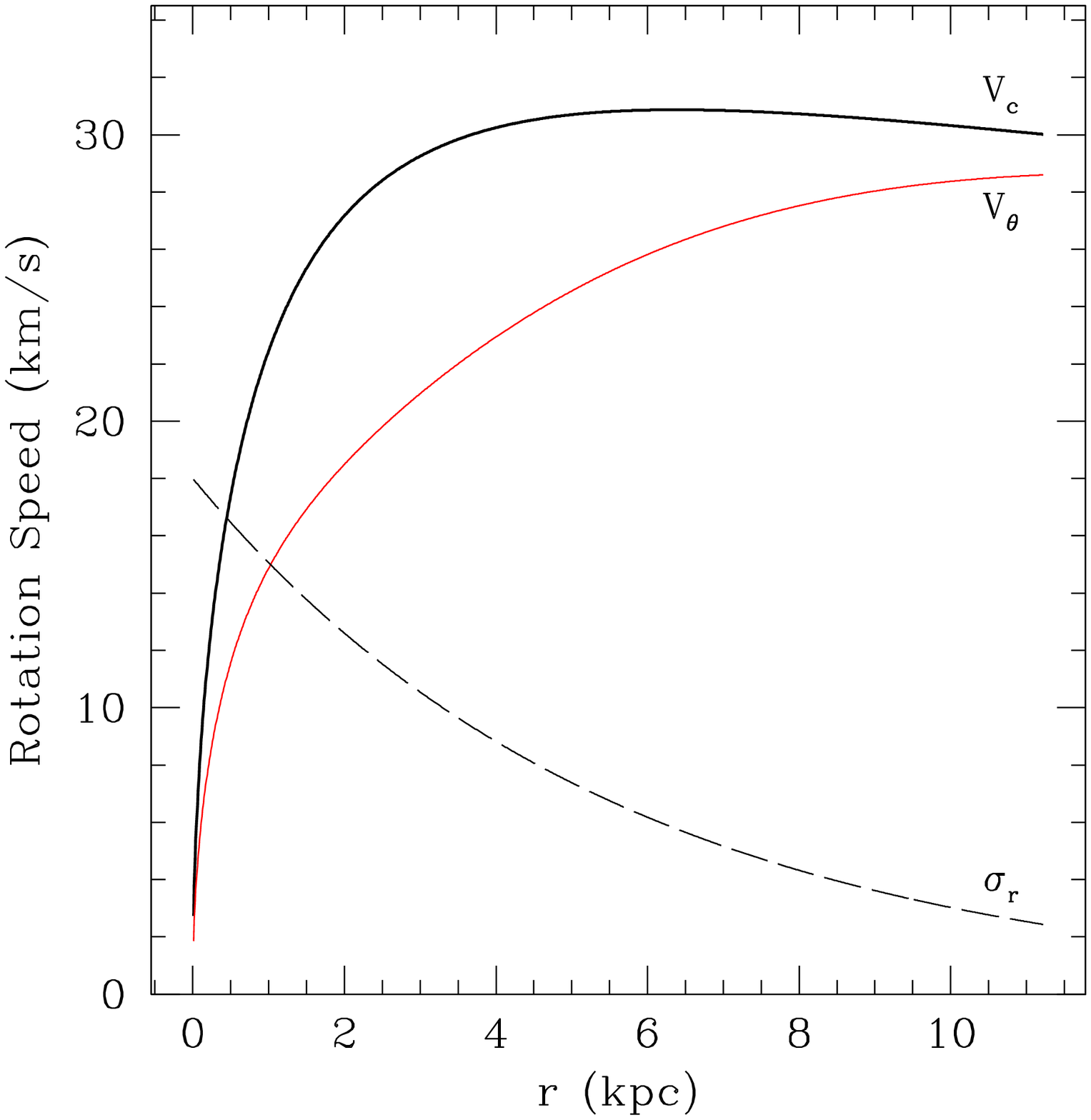}
\includegraphics[width=2.25in]{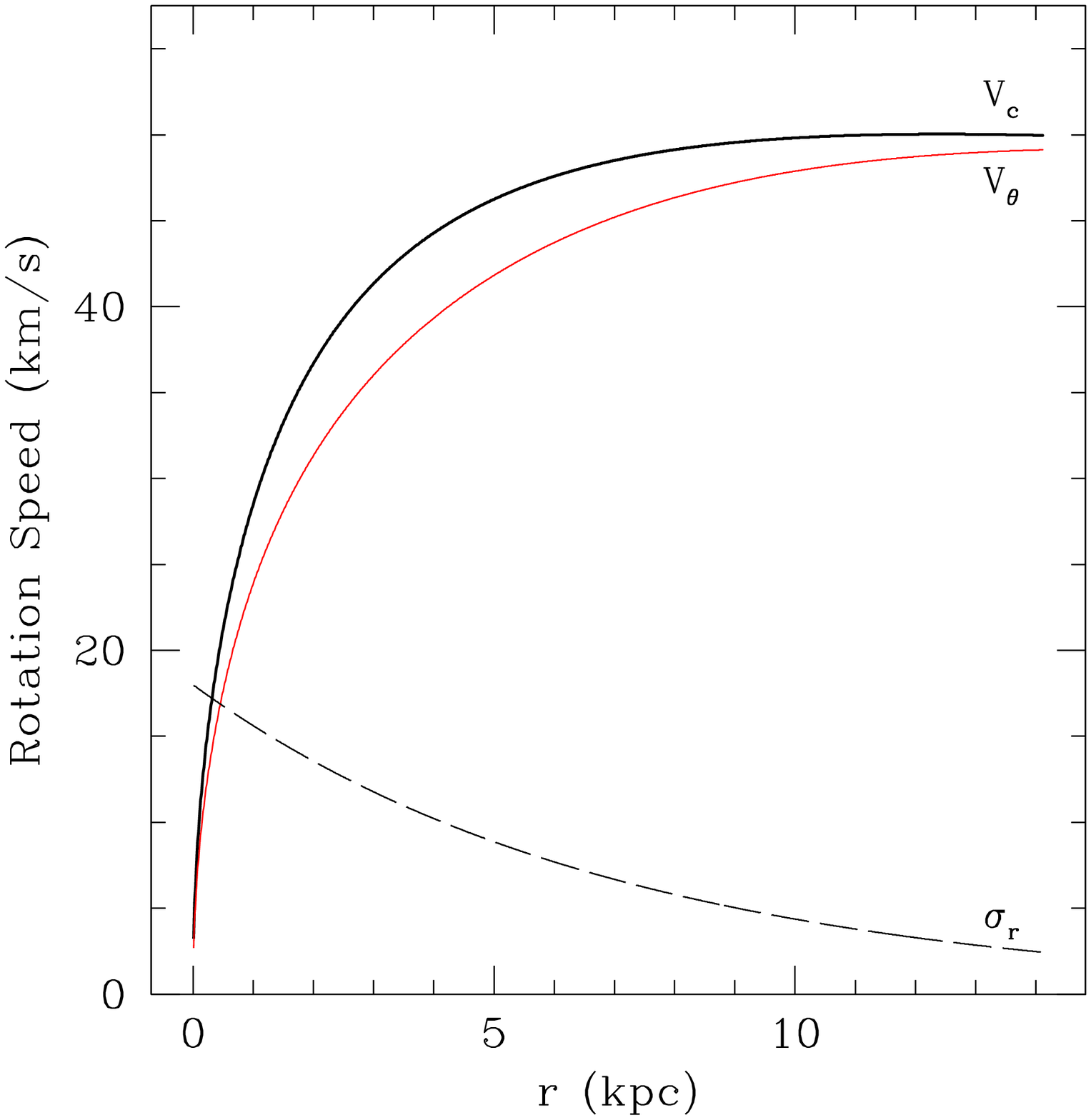}
\includegraphics[width=2.25in]{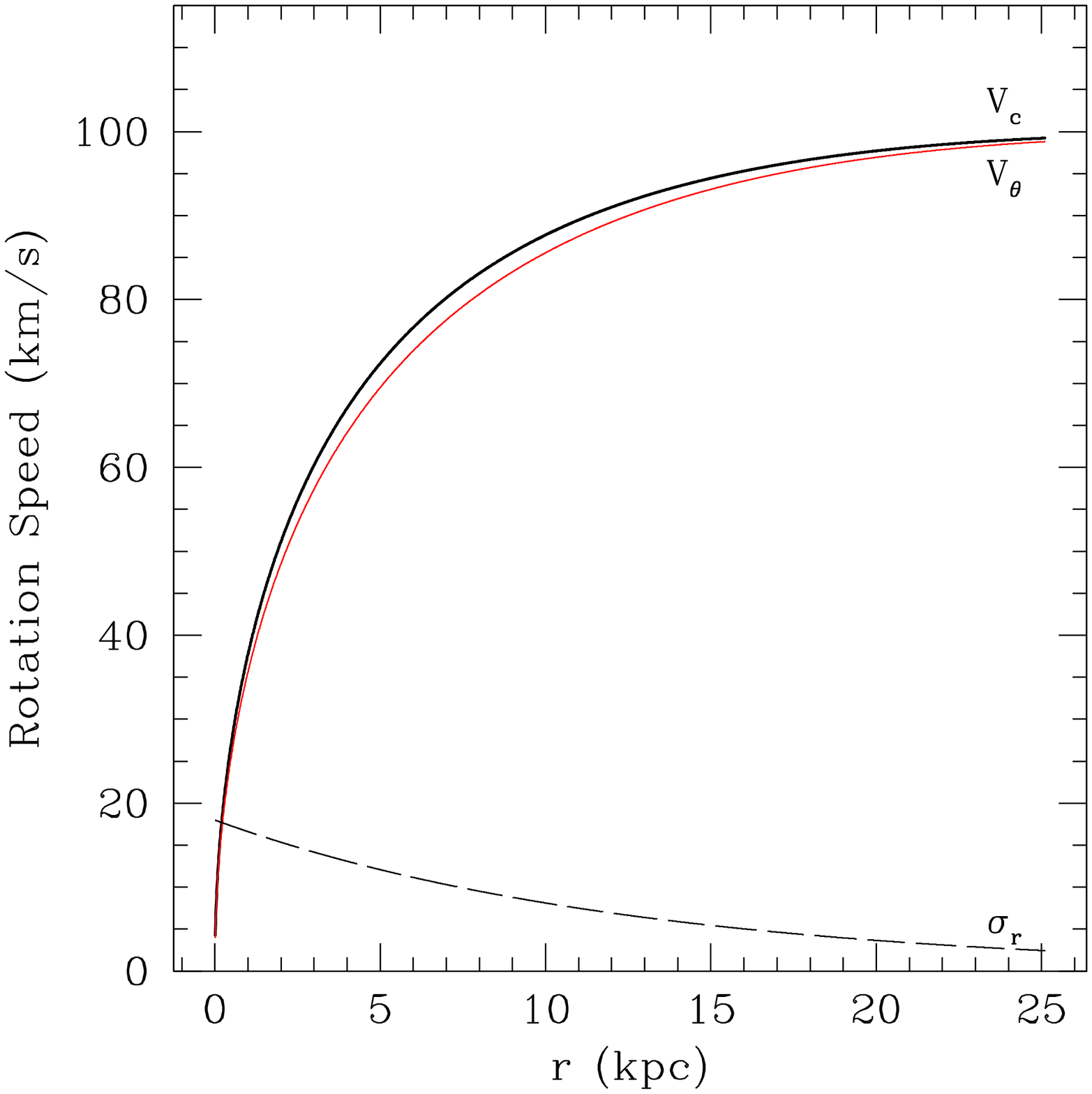}
}
\caption{True (heavy black; $V_c$) and apparent (light red;
  $V_\theta$) circular speed, including the effects of pressure
  support, for a fiducial intrinsic NFW density profile in the WMAP5
  cosmology, and assuming $h_\sigma/h_r=5$.  The velocity dispersion
  profile is plotted with a dashed line. The difference between the
  true and observed rotation speed is more pronounced for less
  concentrated halos (which have more slowly rising rotation curves),
  and for larger values of $h_\sigma/h_r$ (not
  shown).\label{vcmodelnfwfig}}
\end{figure}

\begin{figure}
\centerline{
\includegraphics[width=2.25in]{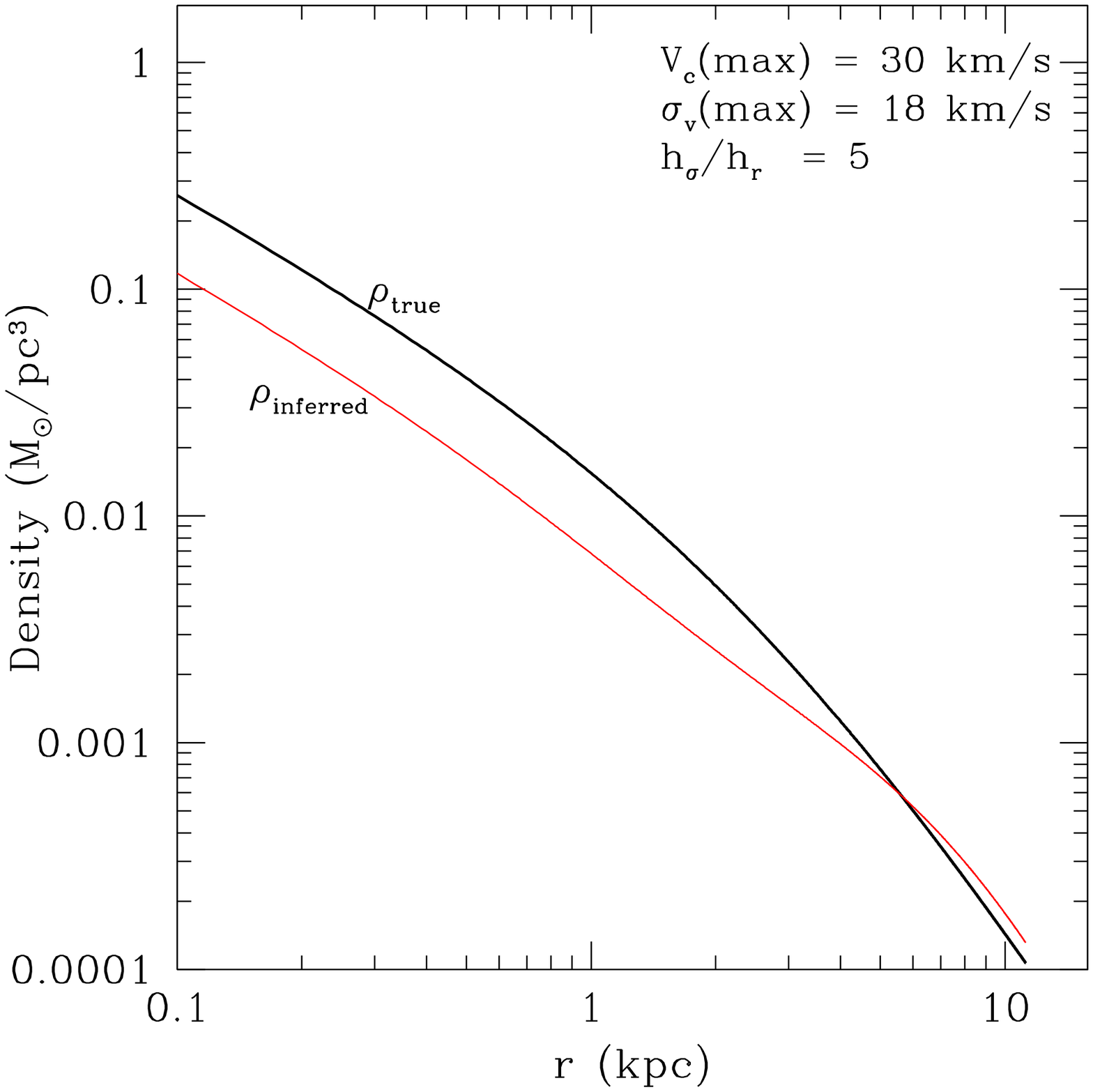}
\includegraphics[width=2.25in]{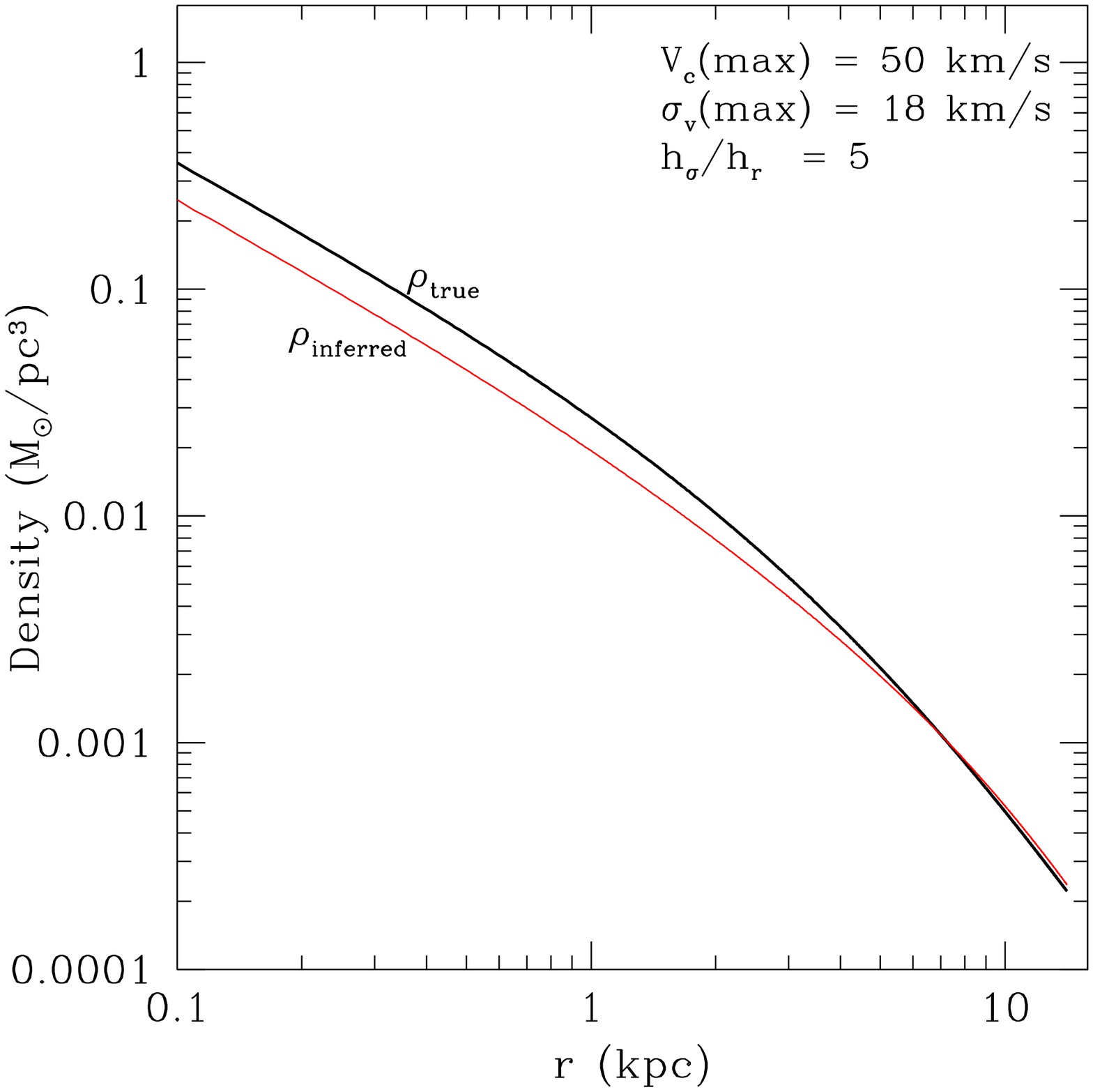}
\includegraphics[width=2.25in]{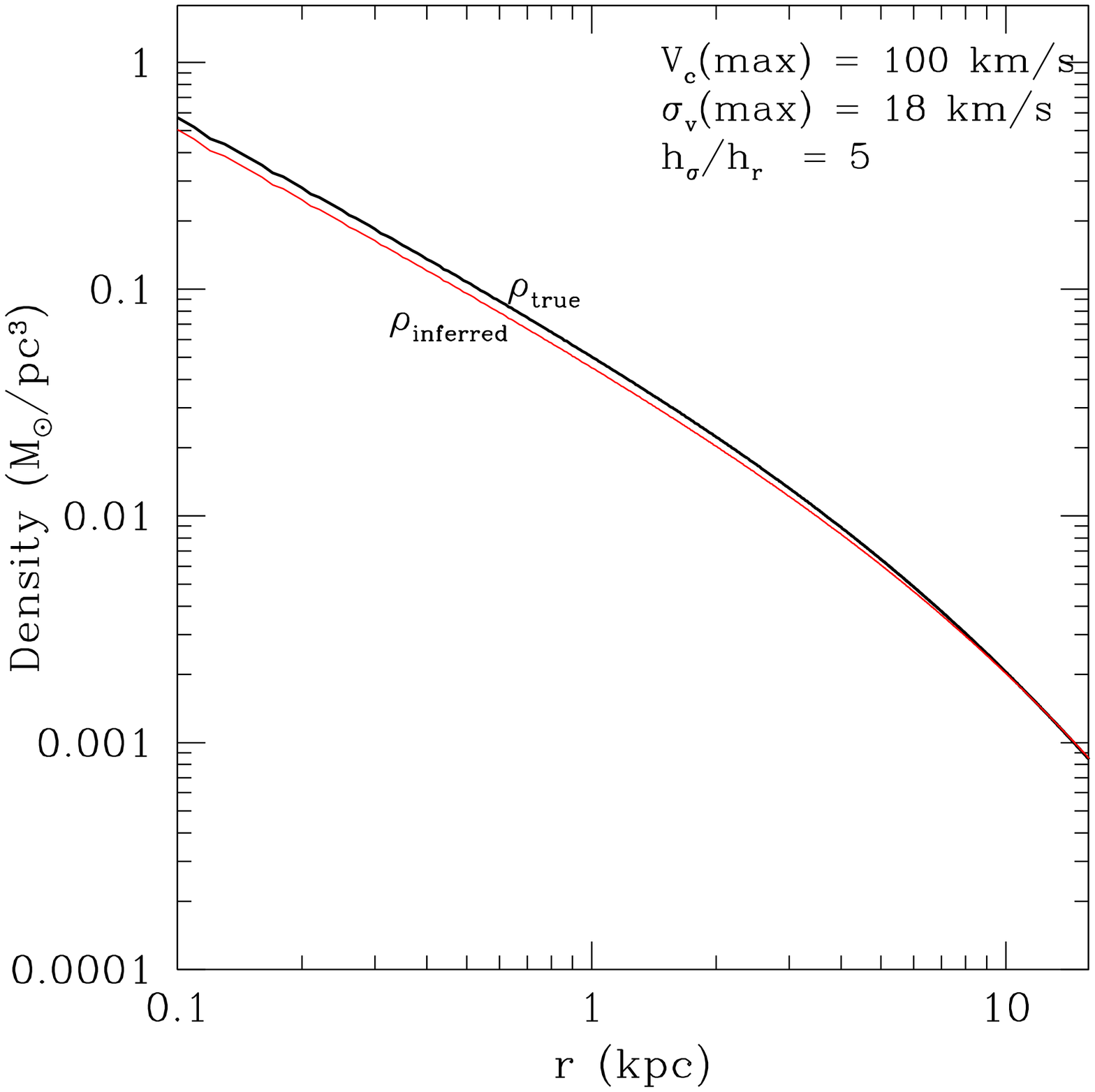}
}
\caption{True (heavy black; $\rho_{true}$) and inferred (light red;
  $\rho_{inferred}$) density profile, including the effects of
  pressure support, for a fiducial intrinsic NFW density profile in a
  WMAP5 cosmology, and assuming $h_\sigma/h_r=5$. The difference
  between the true and observed density is more pronounced for less
  concentrated halos, and for larger values of $h_\sigma/h_r$ (not
  shown).\label{rhomodelnfwfig}}
\end{figure}
\vfill
\clearpage

\begin{figure}[t]
\centerline{
\includegraphics[width=2.25in]{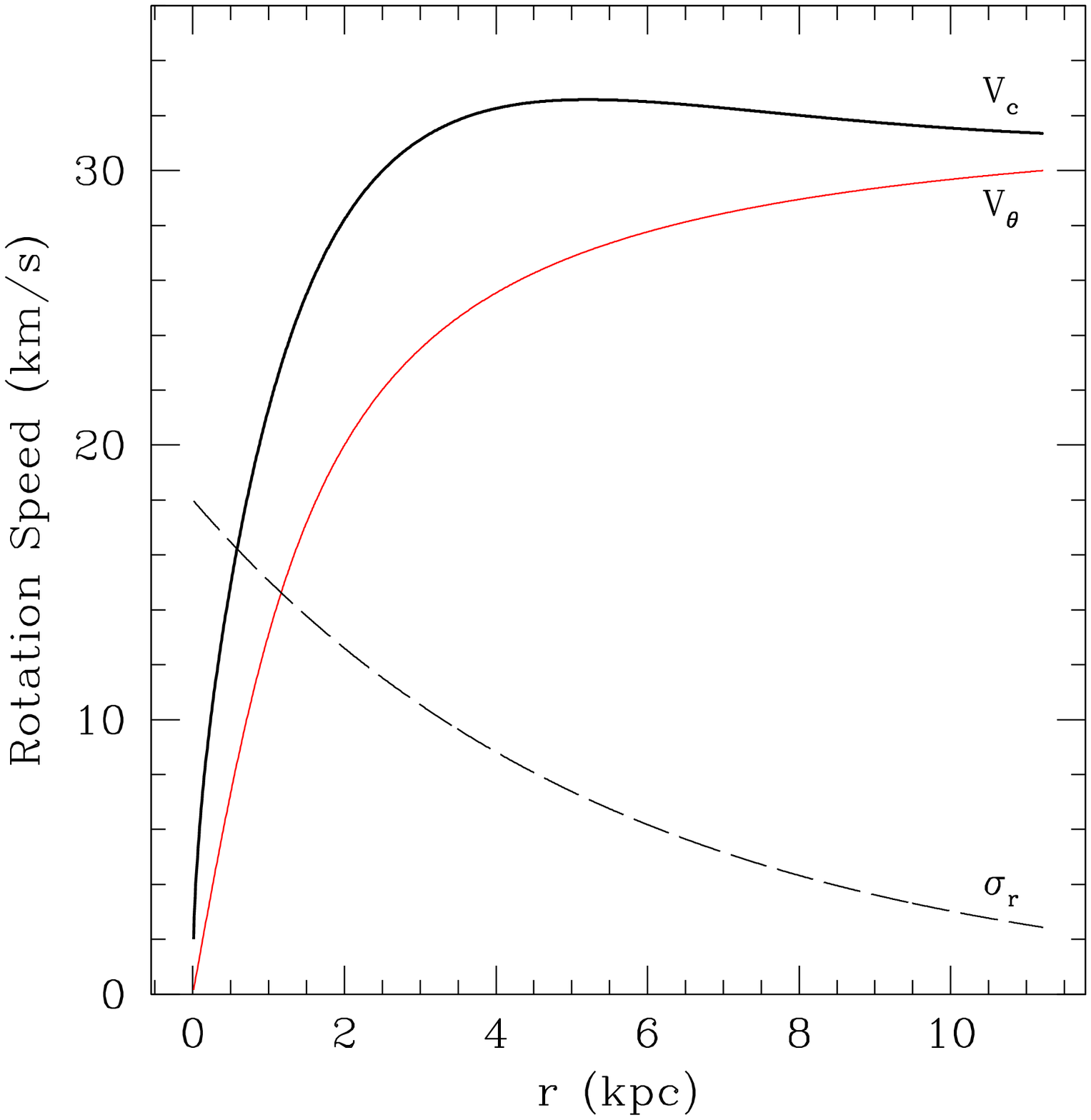}
\includegraphics[width=2.25in]{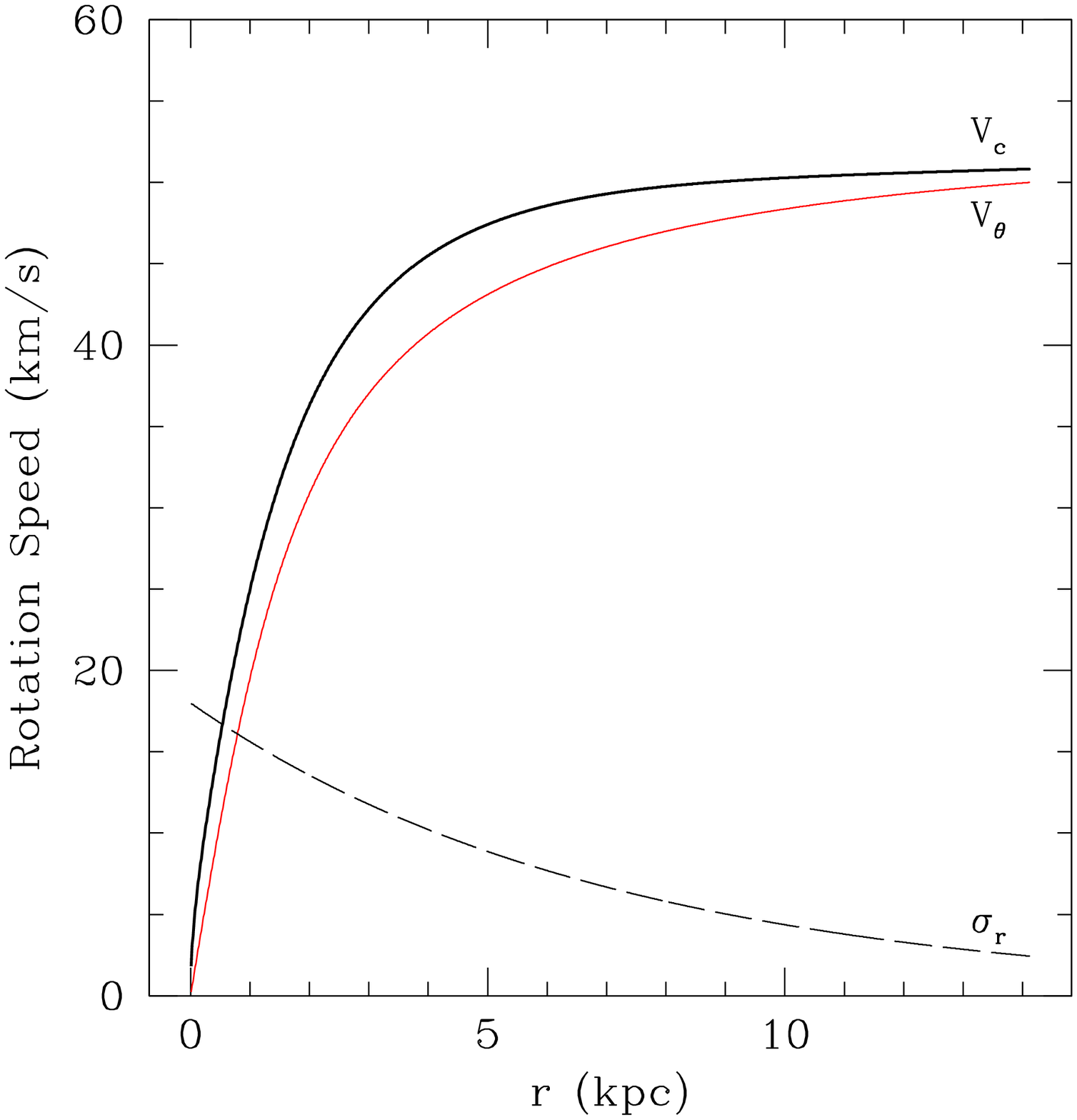}
\includegraphics[width=2.25in]{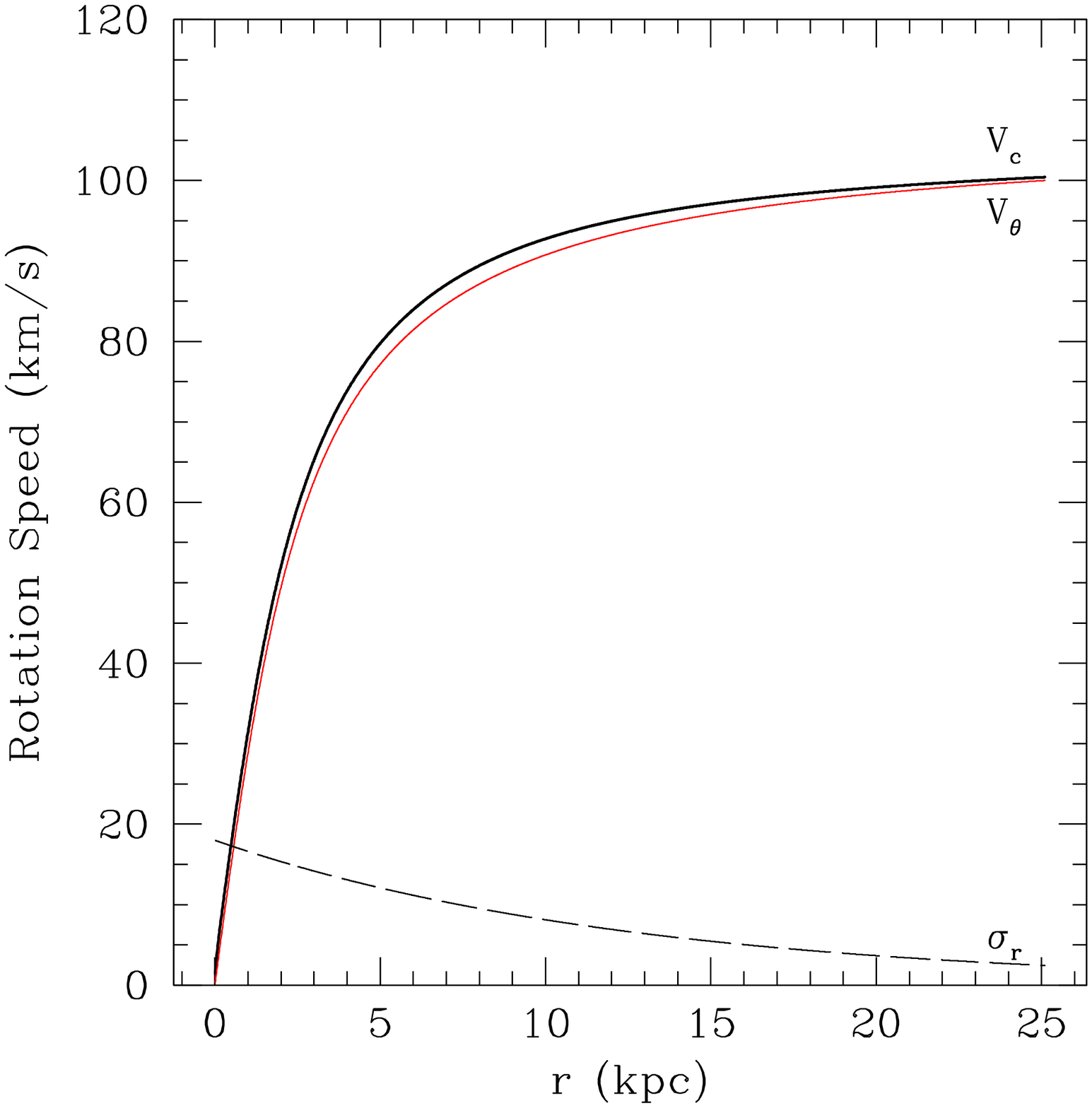}
}
\caption{True (heavy black; $V_c$) and apparent (light red;
  $V_\theta$) circular speed, including the effects of pressure
  support, for an apparent arctan rotation curve, assuming
  $h_\sigma/h_r=5$.  The velocity dispersion profile is plotted with a
  dashed line. The difference between the true and observed rotation
  speed is more pronounced for larger values of $h_\sigma/h_r$, and
  for larger values of $r_t$ at fixed $V_{max}$ (not
  shown).\label{vcmodelisofig}}
\end{figure}

\begin{figure}
\centerline{
\includegraphics[width=2.25in]{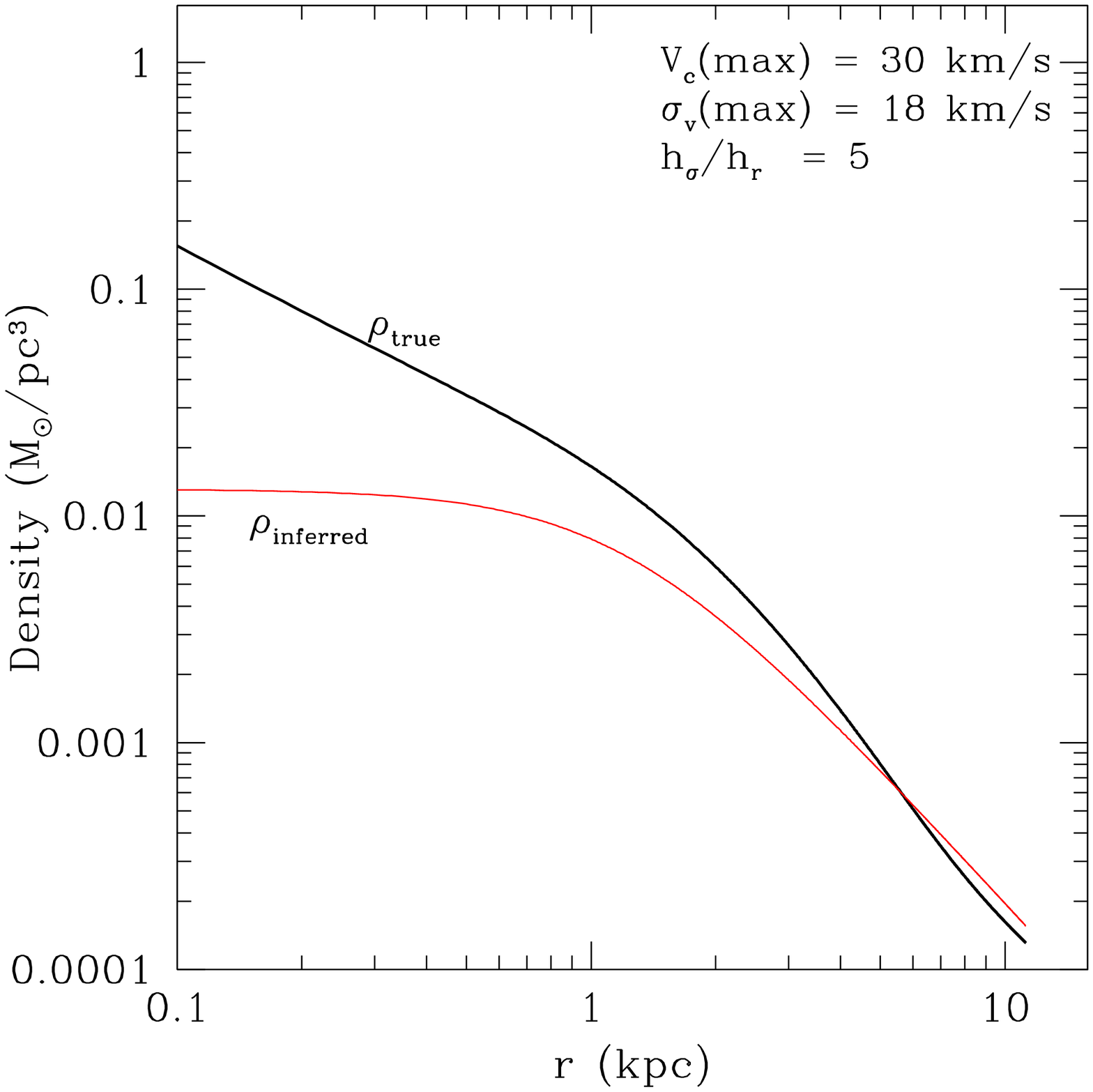}
\includegraphics[width=2.25in]{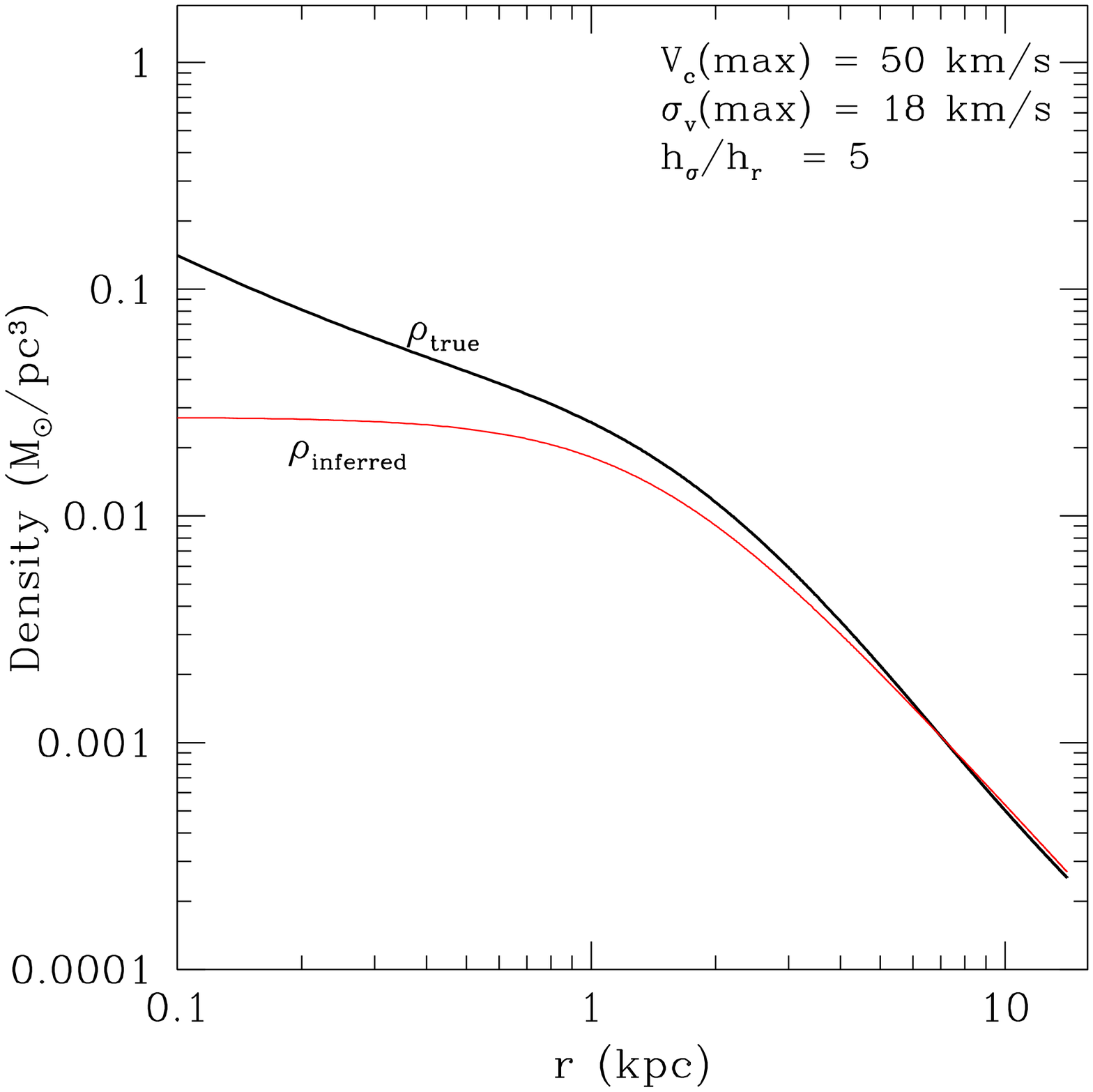}
\includegraphics[width=2.25in]{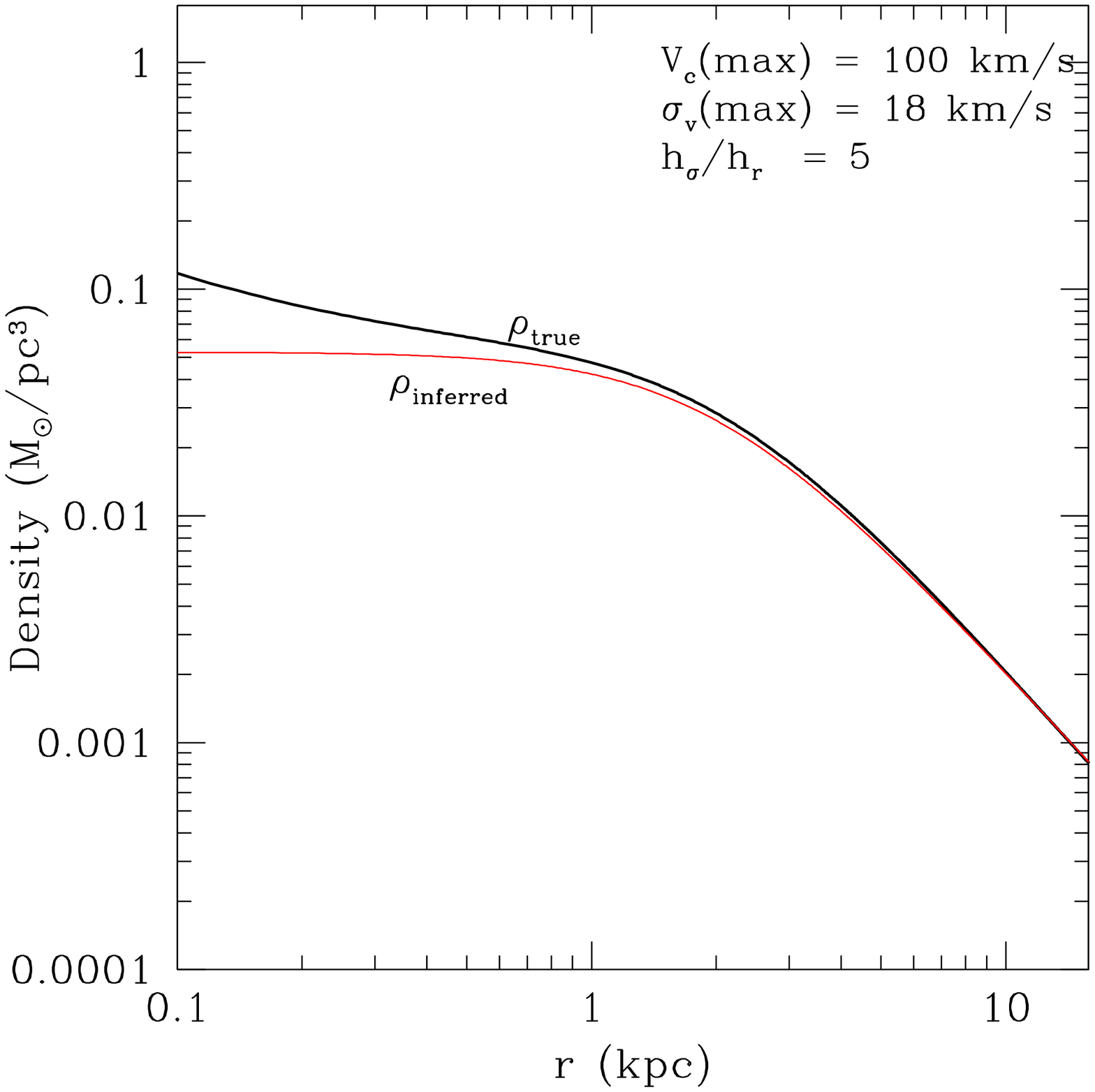}
}
\caption{True (heavy black; $\rho_{true}$) and inferred (light red;
  $\rho_{inferred}$) density profile, including the effects of
  pressure support, assuming $h_\sigma/h_r=5$; for this case, the
  observed rotation curve was fixed to follow an arctan function,
  consistent with a constant density core.  Although the ``observed''
  density profile (i.e.\ what one would infer from the observed arctan
  rotation curve) has a flat inner core, the true density profile
  (inferred after corrections for pressure support) has a central
  cusp, with an inner power law slope of -0.96, -0.7, and -0.35 for
  the $30\kms$, $50\kms$, and $100\kms$ halos, respectively.  The
  inner central slope becomes slightly steeper for larger values of
  $r_t$ or of $h_\sigma/h_r$.\label{rhomodelisofig}}
\end{figure}
\vfill
\clearpage

\begin{figure}[t]
\centerline{
\includegraphics[width=6.5in]{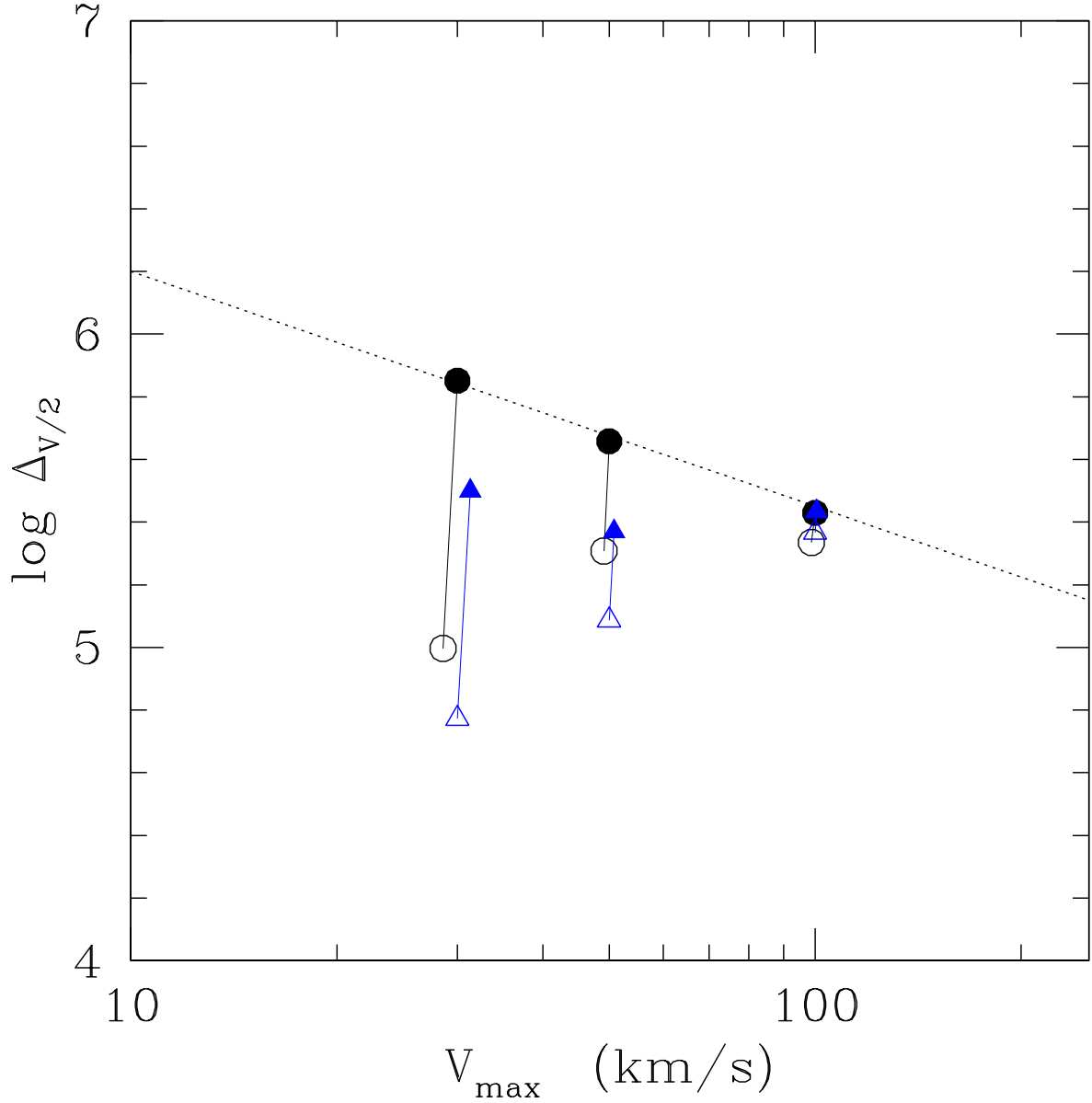}
}
\caption{Dimensionless density as a function of rotation speed, for
  observed (open) and true (filled) rotation curves.  Circles are for
  rotation curves based on true NFW density profiles
  (Figure~\ref{vcmodelnfwfig}), and triangles are for observed
  arctangent rotation curves (Figure~\ref{vcmodelisofig}). The dotted
  line is the approximate relationship for WMAP5, from
  \citet{maccio2008}, for reference.  Failure to correct for pressure
  support can lead to a factor of ten underestimate in the densities
  of low mass dwarf galaxies.  In contrast, the inferred densities of
  higher mass galaxies ($V_c\!\sim\!100\kms$) are unaffected by
  pressure support for our fiducial models.  \label{alamfig}}
\end{figure}
\vfill
\clearpage

\clearpage

\begin{figure}[t]
\centerline{
\includegraphics[width=3.2in]{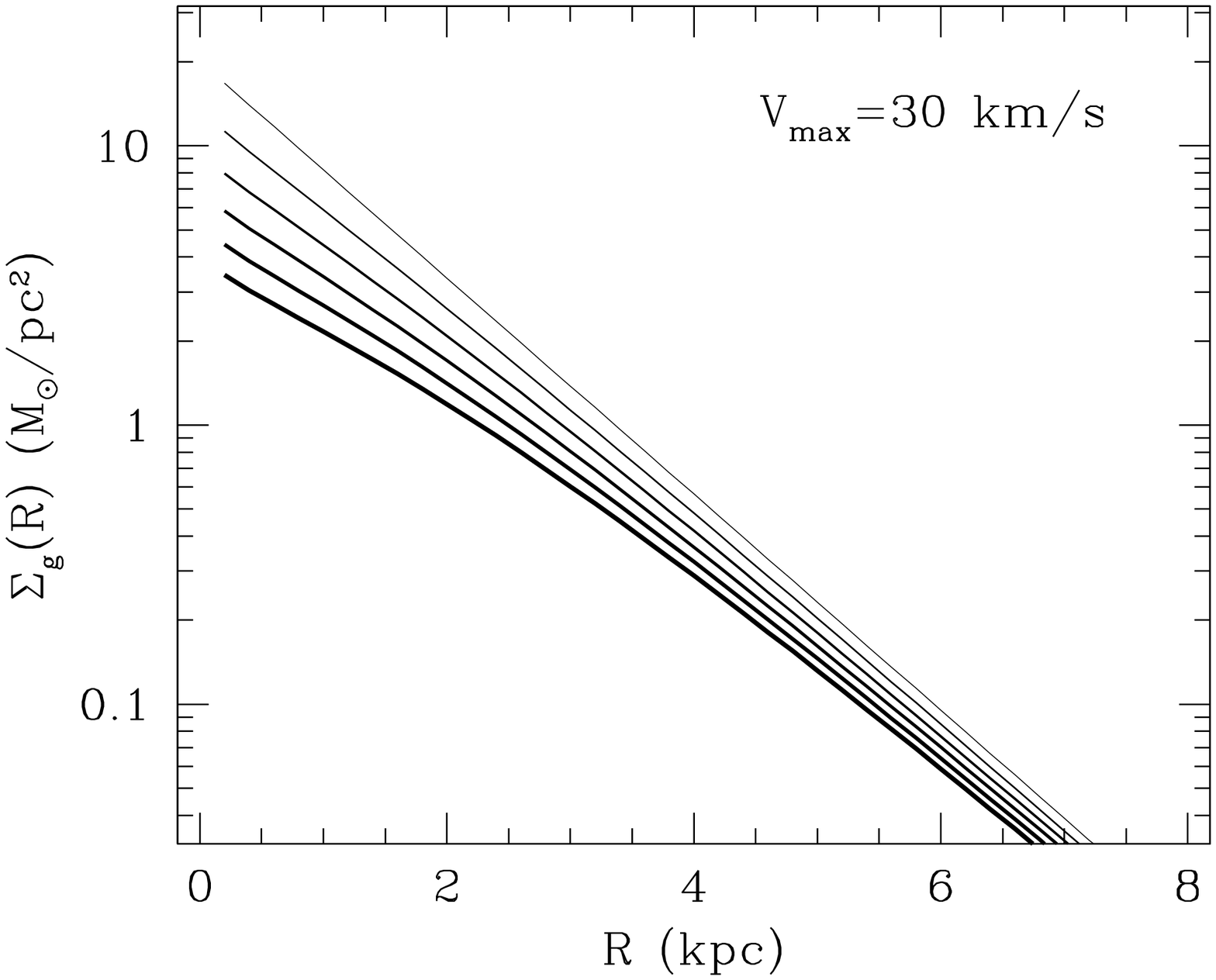}
\includegraphics[width=3.2in]{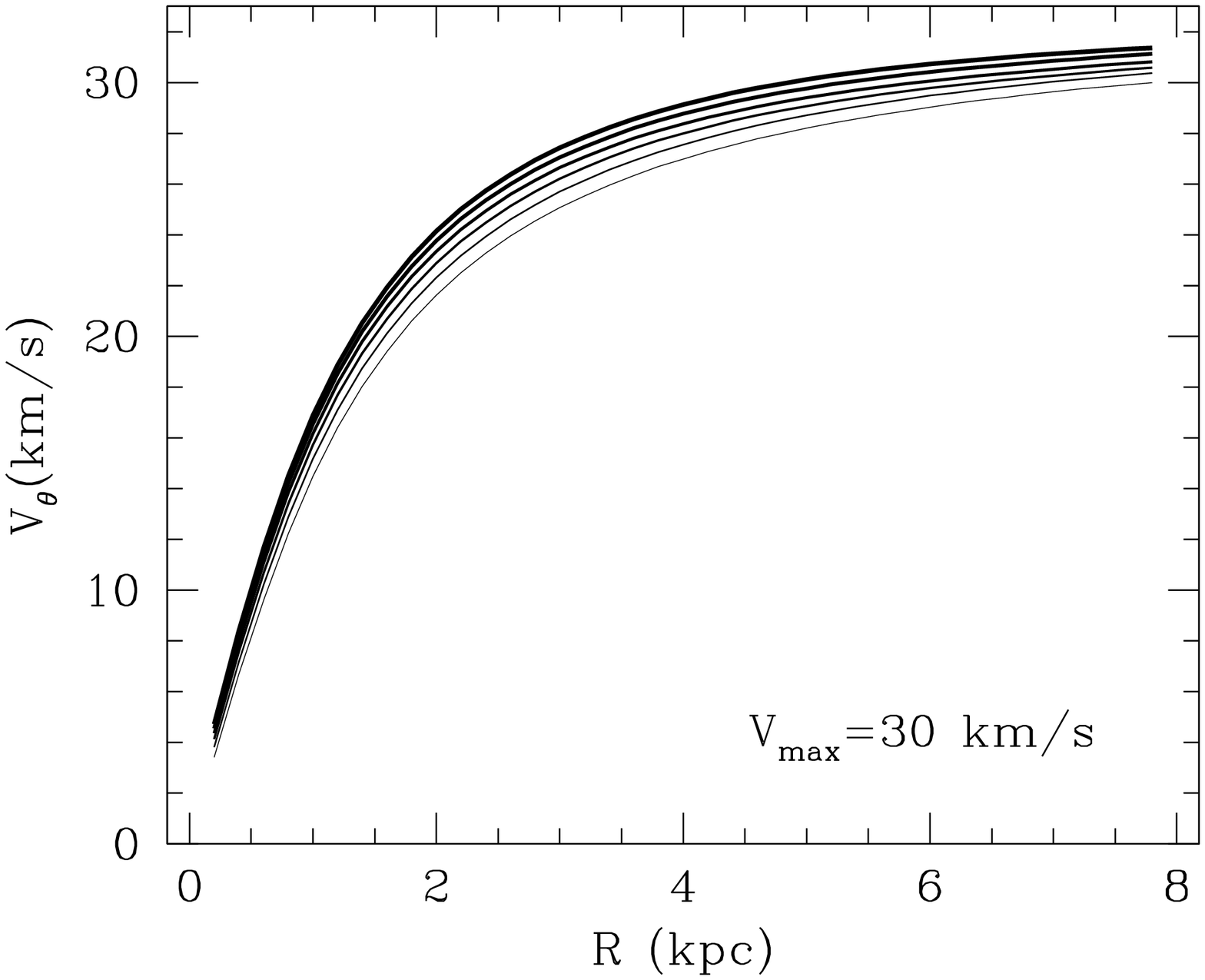}
}
\centerline{
\includegraphics[width=3.2in]{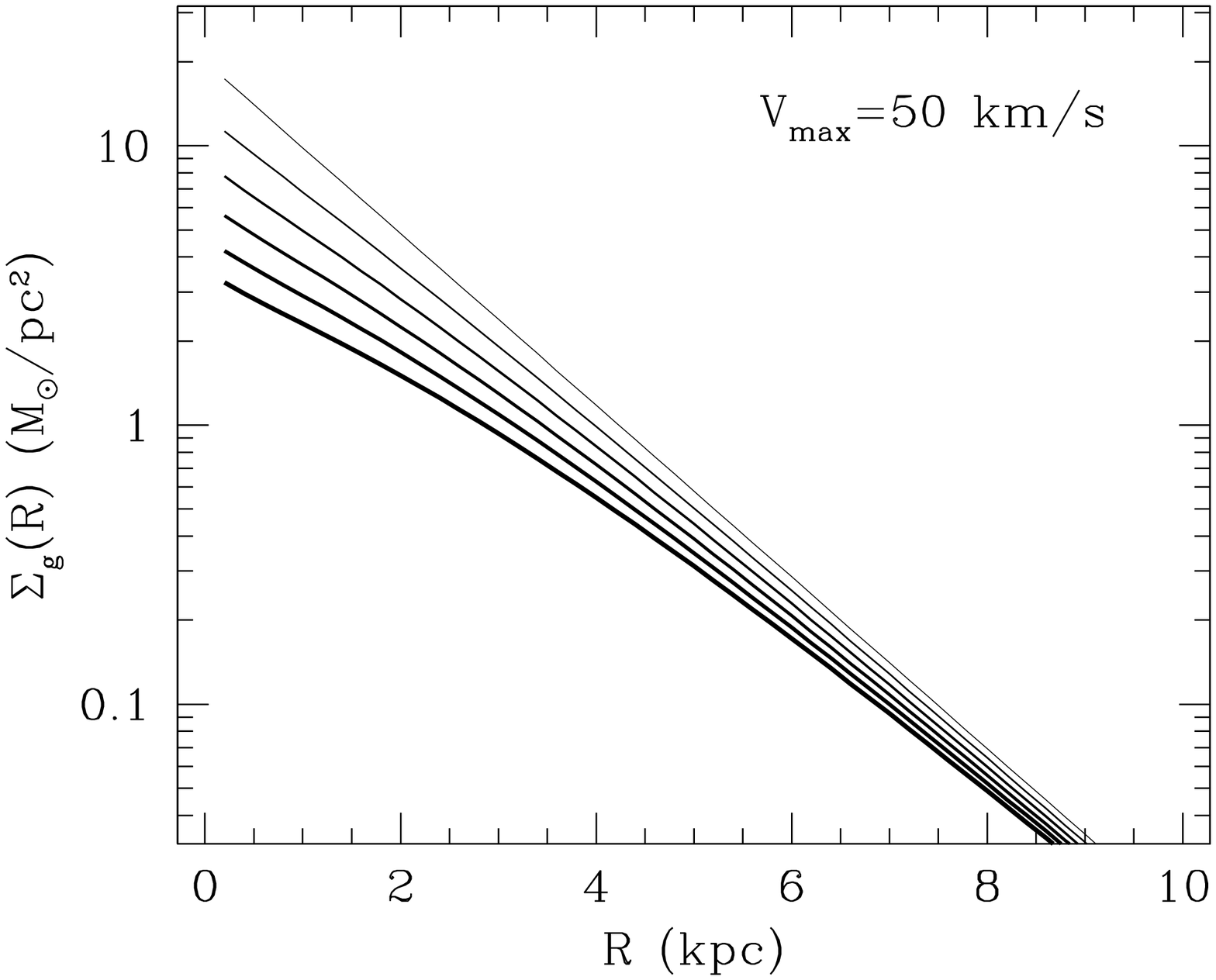}
\includegraphics[width=3.2in]{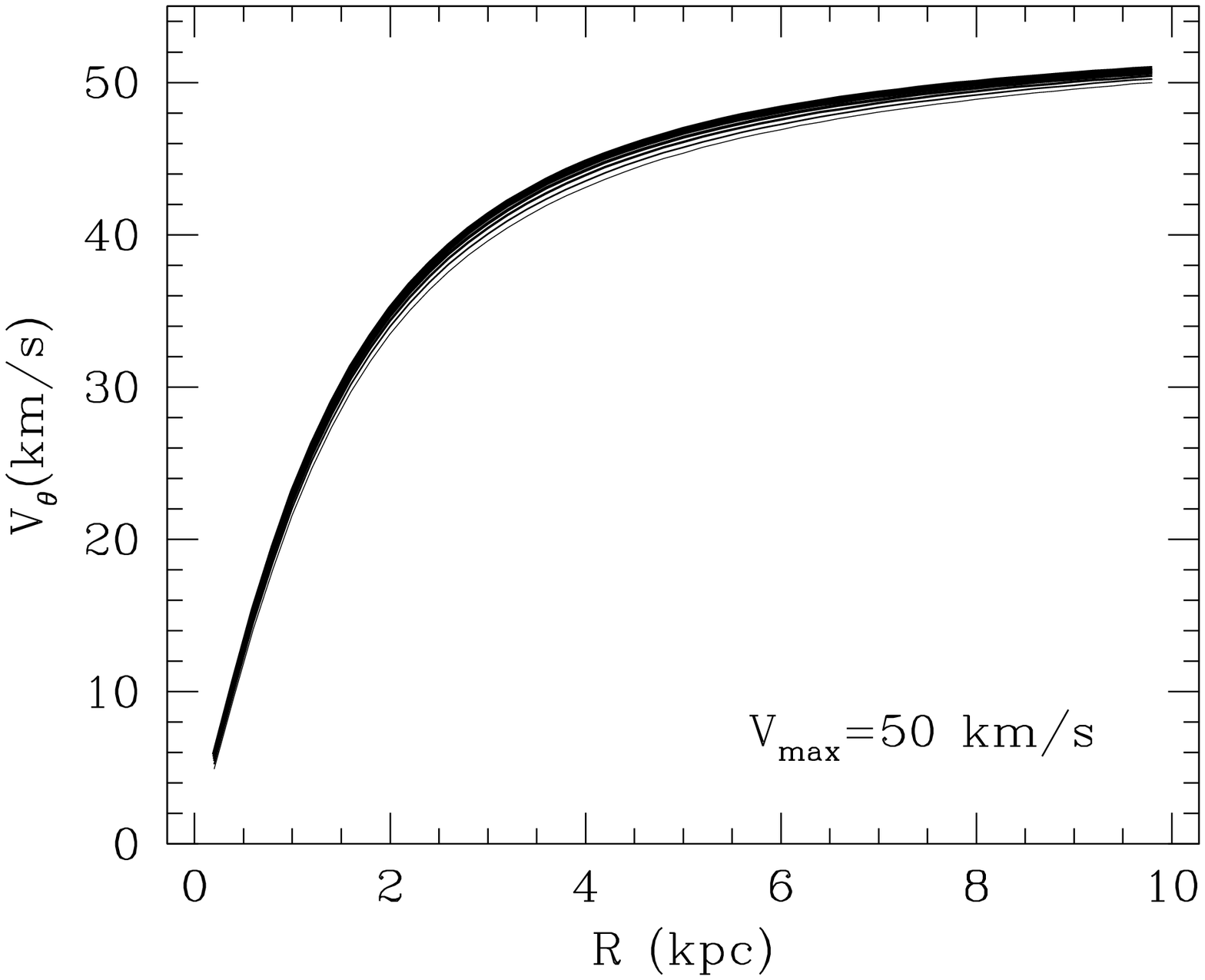}
}
\centerline{
\includegraphics[width=3.2in]{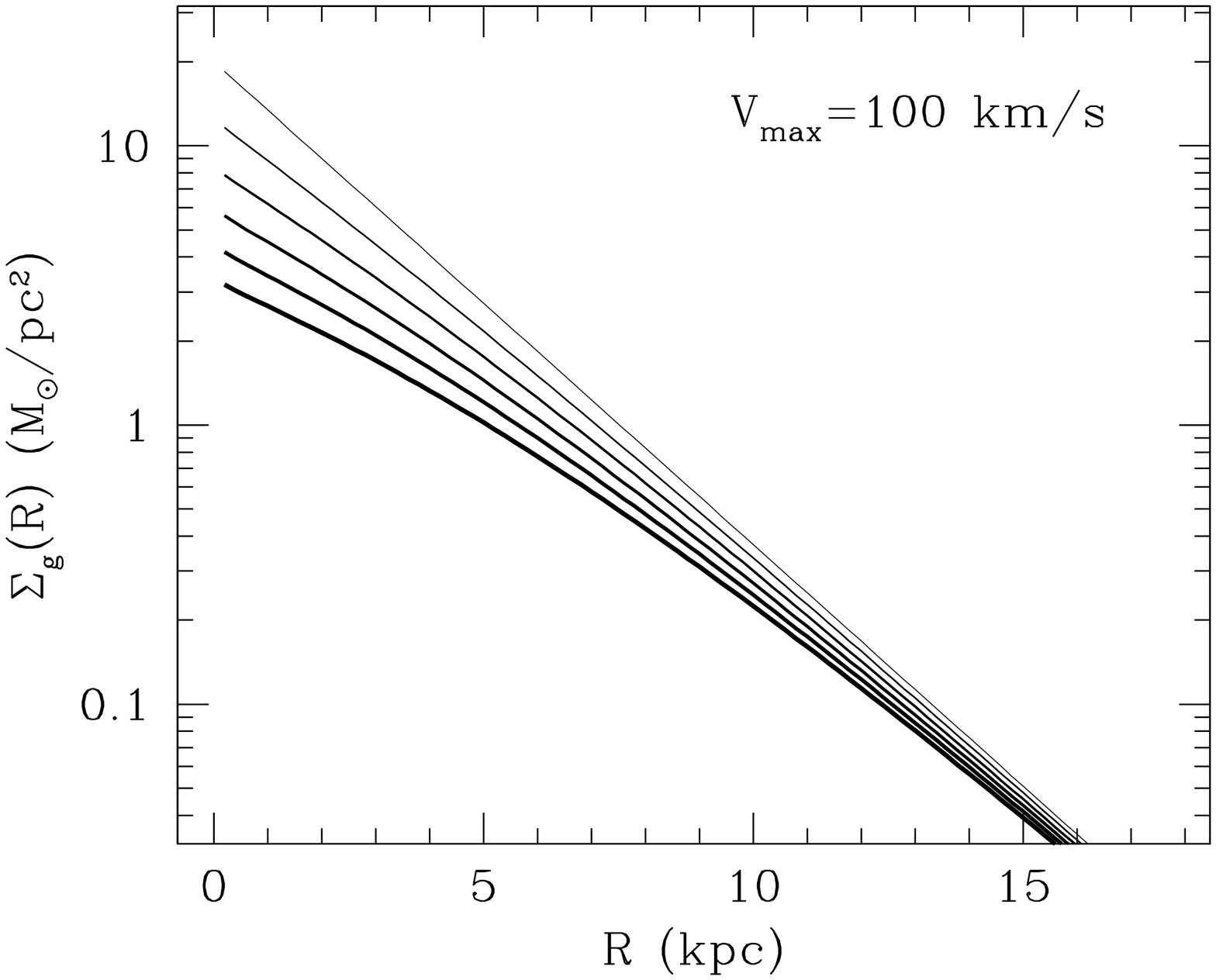}
\includegraphics[width=3.2in]{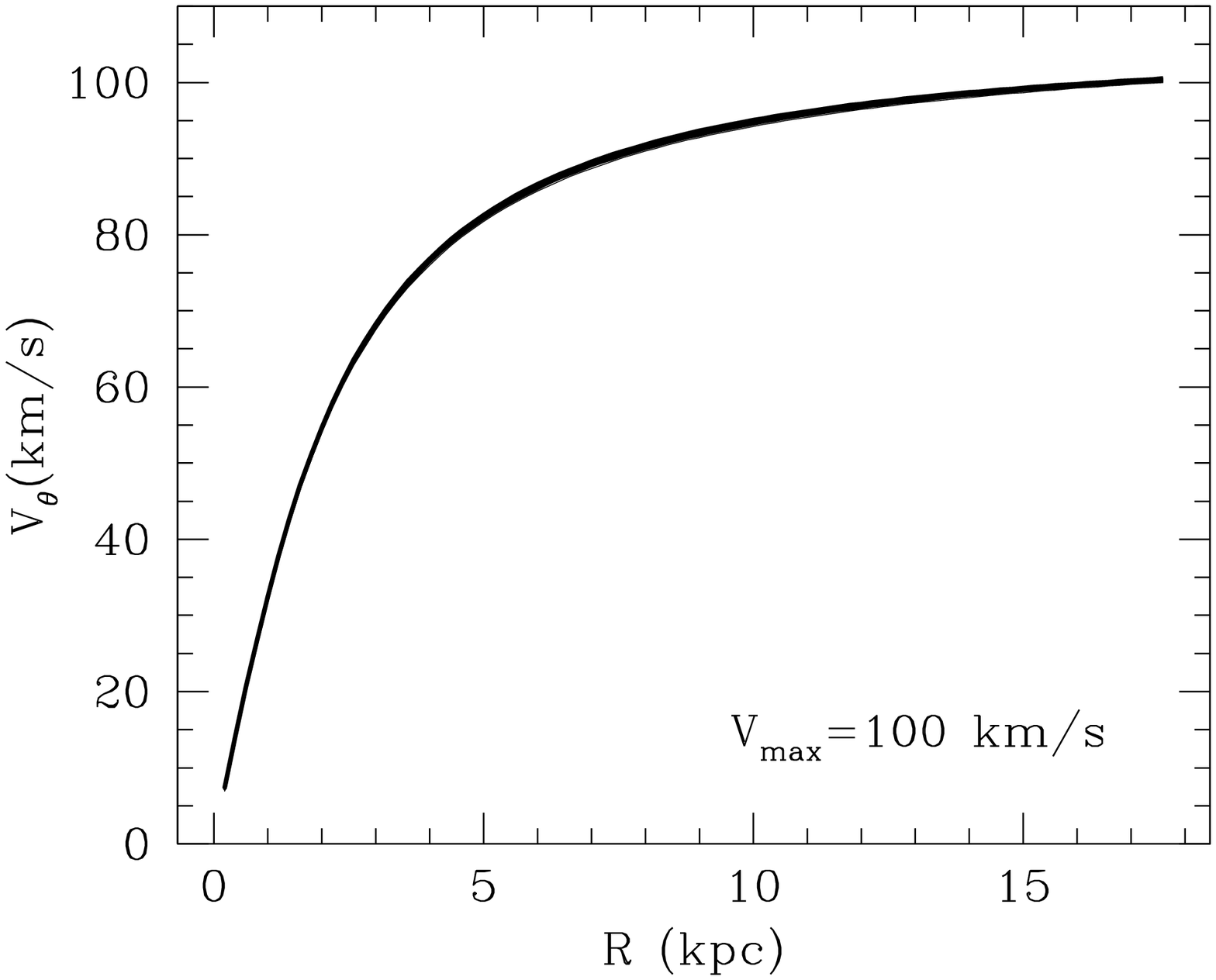}
}
\caption{Evolution of gas surface density (left) and rotation speed
  (right) in rotating disks, including the effects of pressure
  support.  The gas disk starts with an exponential surface density
  (light line), and evolves due to gas consumption and radial
  redistribution in response to declining pressure support.
  Successively darker lines show the evolution in 1\Gyr timesteps, for
  fiducial halos with $V_{max}=30$,\,50,\,\&\,$100\kms$ (top to
  bottom) and a constant turbulent velocity of $\sigma_r=15\kms$.
  Evolution in the gas density due to star formation does not
  drastically alter the degree of pressure support in the disk, due to
  the resulting radial inflow of gas from the outer disk.  \label{diskevfig}}
\end{figure}
\vfill

\clearpage

\begin{figure}[t]
\centerline{
\includegraphics[width=3.3in]{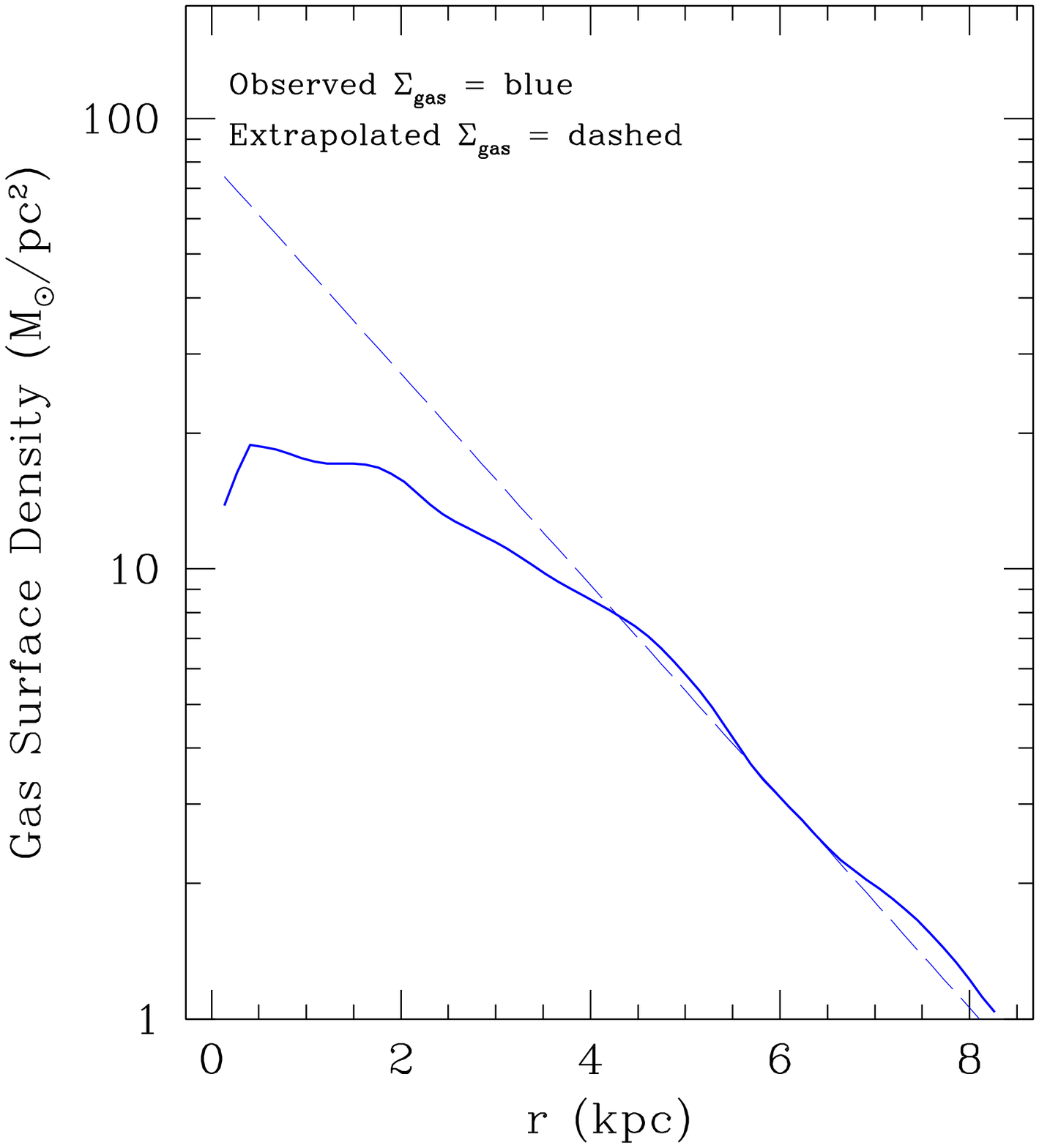}
\includegraphics[width=3.3in]{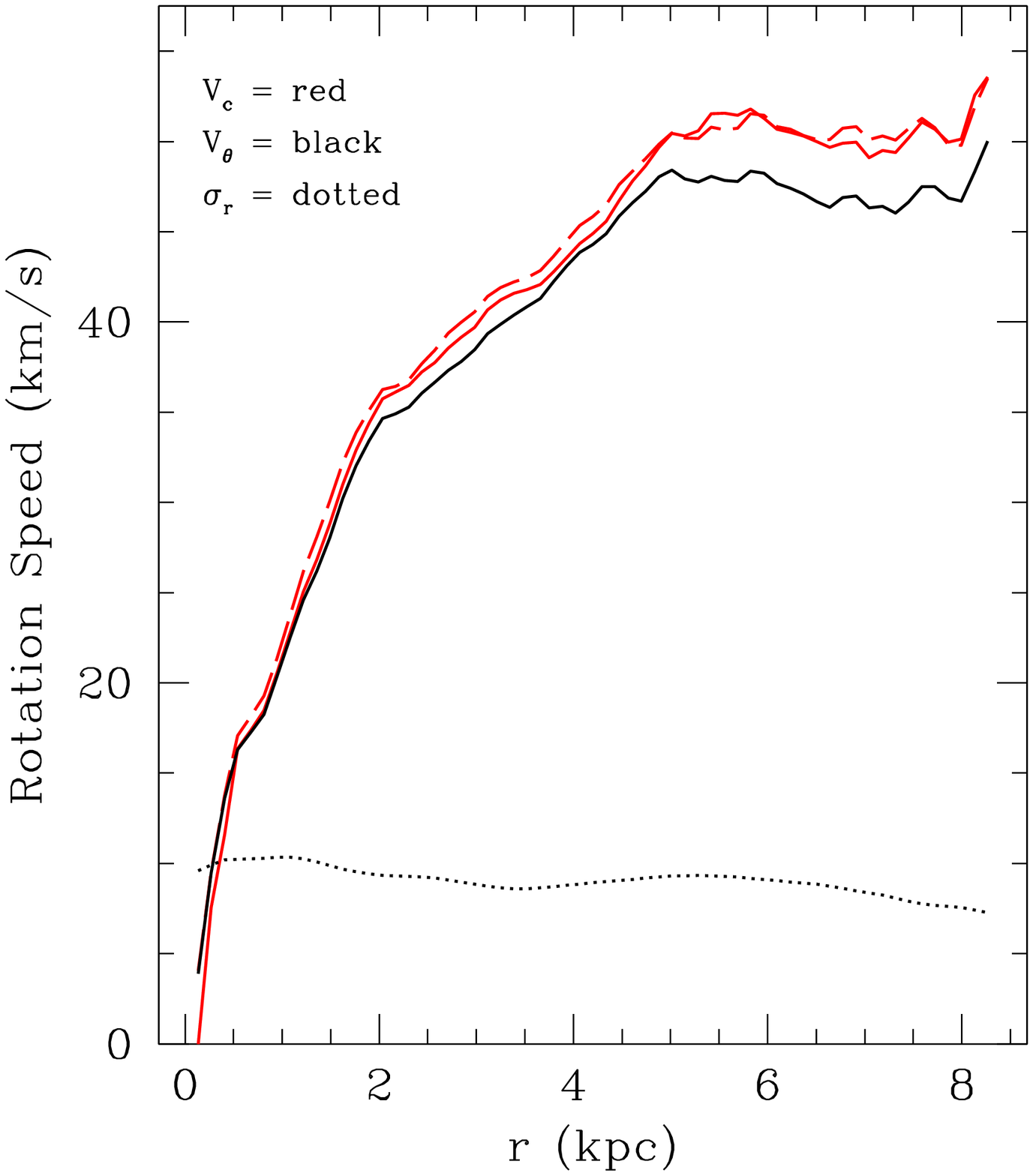}
}
\caption{Gas surface density (left) and velocity profiles (right) for
  DDO~154, derived from H{\sc i} data from the THINGS survey.  The
  left hand panel shows the inferred gas surface density profile,
  assuming that $\Sigma_{gas}=1.4\times\Sigma_{HI}$.  The dashed line
  shows the gas profile if one assumes that the total gas density
  profile is an exponential, with a center dominated by undetected
  molecular gas.  The right hand panel shows the observed H{\sc i}
  line-of-sight velocity dispersion profile (dotted line), and the
  observed tangential velocity (solid black line), corrected for
  inclination.  The red lines show the effects of correcting for
  pressure support assuming an exponential gas distribution (dashed
  line) and the observed gas distribution (solid line).  Pressure
  support appears to be negligible in the inner regions of this
  particular galaxy, as expected based on the models in
  Figure~\ref{rhomodelisofig}.  \label{ddo154fig}}
\end{figure}
\vfill

\fi

\end{document}